\documentstyle[epsf,epsfig,rotate]{mn2e}

\newcommand{\nab}{\mbox{\boldmath $\nabla$}}
\newcommand{\tim}{\mbox{\boldmath $\times$}}
\newcommand{\xib}{\mbox{\boldmath $\xi$}}
\newcommand{\beq}{ \begin{equation} }
\newcommand{\eeq}{ \end{equation} }

\def\spose#1{\hbox to 0pt{#1\hss}}
\def\ltsim{\mathrel{\spose{\lower.5ex\hbox{$\mathchar"218$}}
         \raise.4ex\hbox{$\mathchar"13C$}}}

\onecolumn

\title[Stopping migration by a toroidal magnetic field]
{Stopping inward planetary migration by a toroidal magnetic field}

\author[Caroline Terquem] {Caroline E. J. M. L. J. Terquem \\ 
Institut d'Astrophysique de Paris, 98 bis Boulevard Arago, 75014
Paris, France ---  terquem@iap.fr \\
Universit\'e Denis Diderot--Paris VII, 2 Place Jussieu, 75251
Paris Cedex 5, France}

\date{Accepted.
      Received;
      in original form }

\pubyear{}

\begin{document} 

\maketitle

\begin{abstract} 
We calculate the linear torque exerted by a planet on a circular orbit
on a disc containing a toroidal magnetic field.  All fluid
perturbations are singular at the so--called {\em magnetic
resonances}, where the Doppler shifted frequency of the perturbation
matches that of a slow MHD wave propagating along the field line.
These lie on both sides of the corotation radius.  Waves propagate
outside the Lindblad resonances, and also in a restricted region
around the magnetic resonances.

The magnetic resonances contribute to a significant global torque
which, like the Lindblad torque, is negative (positive) inside
(outside) the planet's orbit.  Since these resonances are closer to
the planet than the Lindblad resonances, the torque they contribute
dominates over the Lindblad torque if the magnetic field is large
enough.  In addition, if $\beta \equiv c^2/v_A^2$ increases fast
enough with radius, the outer magnetic resonance becomes less
important and the total torque is then negative, dominated by the
inner magnetic resonance.  This leads to outward migration of the
planet.  Even for $\beta \sim 100$ at corotation, a negative torque
may be obtained.  A planet migrating inward through a nonmagnetized
region of a disc would then stall when reaching a magnetized region.
It would then be able to grow to become a terrestrial planet or the
core of a giant planet.  In a turbulent magnetized disc in which the
large scale field structure changes sufficiently slowly, a planet may
alternate between inward and outward migration, depending on the
gradients of the field encountered.  Its migration could then become
diffusive, or be limited only to small scales.

\end{abstract}  


\section{Introduction} 

Almost 20\% of the extrasolar planets detected so far orbit at a
distance between 0.038 and 0.1 astronomical unit (au) from their host
star.  It is very unlikely that these short--period giant planets,
also called 'hot Jupiters', have formed {\it in situ \/} (Bodenheimer
1998; Bodenheimer, Hubickyj \& Lissauer 1999).  More likely, they have
formed further away in the protoplanetary nebula and have migrated
down to small orbital distances.  It is also possible that migration
and formation were concurrent (Papaloizou \& Terquem 1999).

So far, three mechanisms have been proposed to explain the location of
planets at very short orbital distances.  One of them relies on the
gravitational interaction between two or more giant planets, which may
lead to orbit crossing and to the ejection of one planet while the
other is left in a smaller orbit (Rasio \& Ford 1996; Weidenschilling
\& Marzari 1996).  However, this mechanism cannot account for the
relatively large number of hot Jupiters observed. Another mechanism is
the so--called 'migration instability' (Murray et al. 1998; Malhotra
1993).  It involves resonant interactions between the planet and
planetesimals located inside its orbit which lead to the ejection of a
fraction of them while simultaneously causing the planet to migrate
inward.  Although such interactions may lead to the migration of
giant planets on small scales, they would require a very massive disc
to move a Jupiter mass planet from several astronomical units down to
very small radii.  Such a massive disc is unlikely and furthermore it
would be only marginally gravitationally stable.  The third mechanism,
that we are going to focus on here, involves the tidal interaction
between the protoplanet and the gas in the surrounding protoplanetary
nebula (Goldreich \& Tremaine 1979, 1980; Lin \& Papaloizou 1979, 1993
and references therein; Papaloizou \& Lin 1984; Ward 1986, 1997).
Here again the protoplanet can move significantly only if there is at
least a comparable mass of gas within a radius comparable to that of
its orbit. However this is not a problem since this amount of gas is
needed anyway in the first place to form the planet.

Note that hot Jupiters can also be produced by the dynamical
relaxation of a population of planets on inclined orbits, formed
through gravitational instabilities of a circumstellar envelope or a
thick disc (Papaloizou \& Terquem 2001).  However, if objects as
heavy as $\tau$--Boo may be produced {\it via} fragmentation, it is
unlikely that lower mass objects would form that way.

A planet embedded in a gaseous disc launches density waves at the
Lindblad resonances (Goldreich \& Tremaine 1979, hereafter GT79).  The
torque exerted by the planet on these waves is responsible for the
exchange of angular momentum between the disc rotation and the
planet's orbital motion.  The interaction between the planet and the
disc inside/outside its orbit leads to a negative/positive torque on
the disc, and therefore to a gain/loss of angular momentum for the
planet.  In the linear regime, because in a (even uniform) Keplerian
disc the outer Lindblad resonances are slightly closer to the planet
than the inner Lindblad resonances, the interaction with the outer
parts of the disc leads to a larger Lindblad torque than that with the
inner parts (Ward 1986, 1997).  Therefore, the net Lindblad torque
exerted by the planet causes it to lose angular momentum and to move
inward relative to the gas (type~I migration).  Note that there is
also a torque exerted by the planet on the material which corotates
with the perturbation.  Its sign depends on the gradient of vortensity
at corotation (GT79).  However, it is usually found to be less
significant than the net Lindblad torque.  The drift timescale for a
1--10~M$_\oplus$ planet undergoing type~I migration in a standard
gaseous disc is typically between $10^4$ and $10^6$~years, shorter
than the disc lifetime or estimated planetary formation timescales.

When the mass of the planet becomes large enough, the perturbation
becomes nonlinear.  Feedback torques from the disc on the planet,
which have to be taken into account in this regime, then opposes the
motion of the planet and stops it altogether when the planet is
massive enough (Ward 1997).  However, density waves still transport
angular momentum outward.  If they are dissipated locally, the angular
momentum they transport is deposited in the disc in the vicinity of
the planet and a gap may be cleared out (Goldreich \& Tremaine 1980;
Lin \& Papaloizou 1979, 1993 and references therein).  The planet is
then locked into the angular momentum transport process of the disc,
and migrates inward at a rate controlled by the disc viscous timescale
(type~II migration).  Here again, the drift timescale is rather short,
being in the range $10^3$--$10^5$~years for standard parameters.

Type~II migration, which occurs for planets with masses at least
comparable to that of Jupiter, can be avoided if after the planet
forms there is not enough mass left in the disc to absorb its angular
momentum.  However, type~I migration appears inevitable as there has
to be enough gas in the disc to push a forming core inward if this
core is to accrete a massive envelope to become a giant planet at some
point.  Different scenarii for stopping a planet undergoing inward
migration have been considered (see, e.g., Terquem, Papaloizou \&
Nelson 2000 and references therein), but none of them can
satisfactorily explain the presence of extrasolar planets with
semi--major axes from a few 0.01~au all the way up to several au.

Recently, Papaloizou (2002) has shown that type~I migration reverses
for reasonable disc models once the eccentricity of the orbit becomes
comparable to the disc aspect ratio.  This is because in that
case the planet spends more time near apocentre, where it is being
speeded up, than near pericentre, where it is being slowed down.
Although the interaction with the disc tends to circularize the orbit
of the planet (Goldreich \& Tremaine 1980), significant eccentricity
may be maintained by gravitational interactions between different
planets forming simultaneously (Papaloizou \& Larwood 2000).

Here we investigate the effect of a magnetic field on planet migration
in the linear regime.  We consider a planet on a circular orbit and,
to keep the problem tractable, restrict ourselves to the case of a
purely toroidal field.  This tends to be the dominant component in
discs as it is produced by the shearing of radial field lines.  As
described above, the torque exerted by a planet on a disc depends
mainly on the location of the radii where the perturbation is in
resonance with the free oscillations of the disc.  We therefore expect
a magnetic field to modify the tidal torque, as it introduces more
degrees of freedom in the disc.

The plan of the paper is as follows.  In sections~\ref{sec:basic},
\ref{sec:equilibrium} and~\ref{sec:potential} we give the basic
equations, describe the equilibrium disc model and give the expression
of the perturbing potential acting on the disc.  In
section~\ref{sec:response} we study the disc response to this
perturbing potential.  We first linearize the equations and derive the
second order differential equation describing the disc response
(\S~\ref{sec:linearization}).  We then carry on a WKB analysis
(\S~\ref{sec:wkb}), which shows that magnetosonic waves propagate
outside the outermost turning points.  In \S~\ref{sec:magneticres} we
study the disc response at the locations where it is singular.  We
call these radii {\it magnetic resonances}.  There are two such
resonances, located on each side of the planet's orbit and within the
Lindblad resonances.  In \S~\ref{sec:turning} we calculate all the
turning points associated with the disc response.  On each side of the
planet there are two or three of them, depending on the value of the
azimuthal number $m$.  One of these turning points co\"{\i}ncide with
the Lindblad resonance.  In contrast to the nonmagnetic case, waves
are found to propagate in a restricted region inside the outermost
turning points, around the magnetic resonances.  We then proceed to
give an expression for the tidal torque in general and in the vicinity
of the magnetic resonances in particular in section~\ref{sec:torque}.
Numerical calculations of the torque exerted by the planet on the disc
are presented in section~\ref{sec:numerics}.  We show that the torque
exerted in the vicinity of the magnetic resonances tends to dominate
the disc response when the magnetic field is large enough.  This
torque, like the Lindblad torque, is negative inside the planet's
orbit and positive outside the orbit.  Therefore, if $\beta \equiv
c^2/v_A^2$ increases fast enough with radius, the outer magnetic
resonance becomes less important (it disappears altogether when there
is no magnetic field outside the planet's orbit) and the total torque
becomes negative, dominated by the inner magnetic resonance.  This
corresponds to a positive torque on the planet, which leads to outward
migration.  Finally, we discuss our results in
section~\ref{sec:discussion}.

\section{Basic equations}

\label{sec:basic}

The disc is described by the equation of motion:

\begin{equation}
\rho \left( \frac{\partial {\bf v}}{\partial t} + \left( {\bf v} \cdot
\nab \right) {\bf v} \right) = - \nab P + {\bf F} - \rho \nab \Psi ,
\label{motion0}
\end{equation}

\noindent the equation of continuity:

\begin{equation}
\frac{\partial \rho}{\partial t} + \nab \cdot
\left( \rho {\bf v} \right) = 0 ,
\label{mass0}
\end{equation}

\noindent and the induction equation in the ideal MHD approximation:

\begin{equation}
\frac{\partial {\bf B}}{\partial t} = 
\nab \tim \left( {\bf v} \tim {\bf B} \right),
\label{induction}
\end{equation}

\noindent where 

\begin{equation}
{\bf F} = \frac{1}{\mu_0} \left( \nab \tim {\bf B} \right) \tim {\bf
B}
\end{equation}

\noindent is the Lorentz force per unit volume, $P$ the pressure,
$\rho$ the mass density, ${\bf v}$ the flow velocity, $\Psi$ the
gravitational potential and ${\bf B}$ the magnetic field ($\mu_0$ is
the permeability of vacuum).  SI units are used throughout the paper.
Here we neglect self gravity so that the gravitational potential
$\Psi$ is assumed to be due to a central mass $M_{\ast}$.

We consider a thin disc, so that the above equations can be averaged
over the disc thickness.  To first order we neglect the
$z$--dependence of ${\bf v}$ and $\Psi$, so that the vertically
averaged equation of motion and continuity are, respectively:

\begin{equation}
\Sigma \left[ \frac{\partial {\bf v}}{\partial t} + 
\left( {\bf v} \cdot \nab \right) 
{\bf v} \right] = - \nab \langle P \rangle + \langle {\bf F}
\rangle  - \Sigma \nab \Psi ,
\label{motion}
\end{equation}

\begin{equation}
\frac{\partial \Sigma}{\partial t} + \nab \cdot
\left( \Sigma {\bf v} \right) = 0 ,
\label{mass}
\end{equation}

\noindent where $\Sigma$ is the surface mass density and the brackets denote 
averaging over the disc thickness.

To close the system of equations, we adopt a barotropic equation of
state:

\begin{equation}
\langle P \rangle = \langle P \rangle \left( \Sigma \right).
\label{state}
\end{equation}

\noindent The sound speed $c$ is then given by:

\begin{equation}
c^2 = \frac{d\langle P \rangle}{d\Sigma}.
\label{csound}
\end{equation}

\section{Equilibrium disc structure}

\label{sec:equilibrium}

We adopt a nonrotating cylindrical polar coordinate system $(r,
\varphi, z)$ with origin at the central mass.  We denote $({\bf e}_r,
{\bf e}_{\varphi}, {\bf e}_z)$ the associated unit vectors.  We
suppose that at equilibrium the disc is axisymmetric and in rotation
around a central mass, so that ${\bf v} = (0, r \Omega(r), 0)$, where
$\Omega$ is the angular velocity.  Furthermore, we assume that the
equilibrium configuration contains only a toroidal magnetic field,
i.e. ${\bf B} = (0, B(r,z), 0)$.  The Lorentz force per unit volume is
then:

\begin{equation}
{\bf F} = \frac{1}{\mu_0} \left( - \frac{B}{r} \frac{\partial \left(r
B \right)}{\partial r} , 0 , -B \frac{\partial B}{\partial z} \right).
\end{equation}

\noindent If we assume reflection with respect to the disc midplane
and that $B$ vanishes at the disc surface, this leads to:

\begin{equation}
\langle {\bf F} \rangle = \frac{1}{\mu_0} \left( -\frac{1}{2 r^2}
\frac{d}{dr} \left(r^2 \langle B^2 \rangle \right), 0 , 0 \right) .
\end{equation}

\section{Perturbing potential}
\label{sec:potential}

We consider a planet of mass $M_p \ll M_{\ast}$ on a circular orbit
with radius $r_p$ and angular velocity $\Omega_p =
\sqrt{GM_{\ast}/r_p^3}$.  At the location $(r, \varphi)$ in the disc,
it exerts the gravitational potential:

\begin{equation}
\Psi'_p (r, \varphi, t) = \Psi'_p (r, \phi) = -
\frac{GM_p}{ \left( r_0^2+r_p^2+r^2 - 2rr_p \cos \phi \right)^{1/2}},
\end{equation}

\noindent where $\phi= \varphi - \Omega_p t$ and we have introduced a
softening length $r_0$.  We now expand $\Psi'_p$ in a Fourier series
with respect to the variable $\phi$:

\begin{equation}
\Psi'_p (r, \phi) = \sum_{m=0}^{\infty} \Psi'_m(r) \cos m \phi,
\label{Psifourier}
\end{equation}

\noindent with 

\begin{displaymath}
\Psi'_{m \ne 0}(r) = \frac{2}{\pi} \int_0^{\pi} \Psi'_p (r, \phi) \; \cos m
\phi \; d \phi,
\end{displaymath}

\noindent and

\begin{displaymath}
\Psi'_0(r) = \frac{1}{\pi} \int_0^{\pi} \Psi'_p (r, \phi) \; d \phi.
\end{displaymath}

Using the generalized Laplace coefficients defined as (Ward 1989;
Korycansky \& Pollack 93, hereafter KP93):

\begin{equation}
b_{1/2}^{m} ( \gamma ) = \frac{2}{\pi} \int_0^{\pi} \frac{\cos m \phi \; d
\phi}{ \left( q^2 + p^2 \gamma^2 - 2 \gamma \cos \phi \right)^{1/2}} ,
\end{equation}

\noindent where:

\begin{displaymath}
{\rm if} \; r<r_p: \; \; \gamma = \frac{r}{r_p}, \; \; p=1, \; \;
q^2=1+\frac{r_0^2}{r_p^2},
\end{displaymath}

\begin{displaymath}
{\rm if} \; r>r_p: \; \; \gamma = \frac{r_p}{r}, \; \; q=1, \; \;
p^2=1+\frac{r_0^2}{r_p^2}, 
\end{displaymath}

\noindent we can rewrite $\Psi'_m$ under the form:

\begin{equation}
\frac{\Psi'_{m \ne 0}}{r_p^2 \Omega_p^2}= - \frac{M_p}{M_\ast} \;
\frac{r_p}{r} \; b_{1/2}^{m} ( \gamma ) \; \; \; {\rm if} \; r>r_p,
\end{equation}

\begin{equation}
\frac{\Psi'_{m \ne 0}}{r_p^2 \Omega_p^2}= - \frac{M_p}{M_\ast} \;
b_{1/2}^{m} ( \gamma ) \; \; \; {\rm if} \; r<r_p,
\end{equation}

\noindent and the expression on the right--hand--side has to be
divided by a factor two if $m=0$.  The subscript 'p' indicates that
the quantities have to be evaluated at $r=r_p$.

\section {The disc response}
\label{sec:response}

\subsection{Linearization of the basic equations}
\label{sec:linearization}

In this paper we consider small perturbations, so that the basic
equations can be linearized.  We denote Eulerian perturbations with a
prime.  Since the equilibrium is axisymmetric and steady, we can
expand each of the perturbed quantities in Fourier series with respect
to the variable $\phi$ and solve separately for each value of $m$.
The general problem may then be reduced to calculating the response of
the disc to the real part of a complex potential of the form
$\Psi'_m(r) {\rm exp} \left[ i \left( m\varphi - \omega t \right)
\right]$, where the amplitude $\Psi'_m$ is real and the frequency
$\omega \equiv m \Omega_p$, and summing up over $m$.  We make all
Eulerian fluid state variable perturbations complex by writing:

\begin{equation}
X'(r, \varphi, t) = \sum_{m=0}^{\infty} X'_m (r) e^{i \left(
m\varphi - \omega t \right)} ,
\end{equation}

\noindent where $X$ is any state variable.  The physical
perturbations will be recovered by taking the real part of these
complex quantities.  We denote $\xib$ the Lagrangian displacement, and
write its $m$--th Fourier component as $\xib_m (r) {\rm exp} \left[ i
\left( m\varphi - \omega t \right) \right]$.  Since here we are only
interested in the perturbations induced by the planet in the plane of
the disc, we take $\xi_z=0$.

Linearization of the induction equation~(\ref{induction}), or the
equivalent statement of the conservation of magnetic flux, leads to an
expression for the perturbed magnetic field in the form (Chandrasekhar
\& Fermi 1953):

\begin{equation}
{\bf B}' = \nab \tim \left( \xib \tim {\bf B} \right).
\label{Bp}
\end{equation}

The corresponding perturbation of the Lorentz force per unit volume is:

\begin{equation}
{\bf F}' = \frac{1}{\mu_0} \left[ \left( \nab \tim {\bf B}' \right)
\tim {\bf B} + \left( \nab \tim {\bf B} \right) \tim {\bf B'} \right].
\label{FMp}
\end{equation}

Use of equations~(\ref{Bp}) and~(\ref{FMp}) with $\xi_z=0$ gives:

\begin{eqnarray}
\langle F'_{mr} \rangle & = & \frac{ \langle B^2 \rangle}{\mu_0 r} 
\left[ r \;
\frac{d^2 \xi_{mr}}{d r^2} + \left( \frac{3 b_1}{2} -1
\right) \frac{d \xi_{mr}}{d r} + \left( \frac{b_2}{2} - m^2
\right) \frac{\xi_{mr}}{r} \right] , \\
\langle F'_{m \varphi} \rangle & = & \frac{imb_1}{2 \mu_0} \; 
\frac{ \langle B^2
\rangle \xi_{mr}}{r^2} , 
\label{Fpertphi} \\
\langle F'_{mz} \rangle & = & 0 ,
\end{eqnarray}

\noindent where we have defined the dimensionless quantities:

\begin{equation}
b_1  \equiv  \frac{d \ln \left( r^2 \langle B^2 \rangle \right)
}{d \ln r} , \; \;
b_2  \equiv  \frac{1}{\langle B^2 \rangle } \; \frac{d}{d r}
\left( r^2 \frac{d \langle B^2 \rangle }{d r} \right) .
\label{bs}
\end{equation}

We now linearize the $r$-- and $\varphi$--components of the equation
of motion~(\ref{motion}) and the equation of continuity~(\ref{mass}).
Using $v'_{mr} = i m \sigma \xi_{mr}$, with $\sigma \equiv \Omega -
\Omega_p$, and equations~(\ref{state}) and~(\ref{csound}), we get:

\begin{equation}
-m^2 \sigma^2 \xi_{mr} - 2 \Omega v'_{m \varphi} = -
 \frac{d}{d r} \left( \Psi'_m + W'_m \right)
 +\frac{1}{\Sigma} \left( - \frac{W'_m}{c^2} \langle F_r \rangle +
 \langle F'_{mr} \rangle \right) ,
\label{ximr}
\end{equation}

\begin{equation}
\sigma v'_{m \varphi} + \frac{\sigma \kappa^2}{2 \Omega} \xi_{mr} = -
\frac{1}{r} \left( \Psi'_m + W'_m \right) - \frac{i}{m \Sigma} \langle
F'_{m \varphi} \rangle,
\label{vphim}
\end{equation}

\begin{equation}
\frac{\sigma W'_m}{c^2} = - \frac{1}{r \Sigma}
\frac{d}{d r} \left( r \sigma \Sigma \xi_{mr} \right) -
\frac{v'_{m \varphi}}{r} ,
\label{Wm}
\end{equation}

\noindent where we have found it convenient to use the radial
component of the Lagrangian displacement $\xi_{mr}$ and the azimuthal
component of the Eulerian perturbed velocity $v'_{m \varphi}$ as
variables.  Here $W'= \Sigma' c^2 / \Sigma$ is the linear perturbation
of the enthalpy and $\kappa$ is the epicyclic frequency, which is
defined by:

\begin{displaymath}
\kappa^2 = \frac{2 \Omega}{r} \frac{d \left( r^2 \Omega \right)}
{dr} .
\end{displaymath}

We can further eliminate $v'_{m \varphi}$ and $W'_m$ to get, after
some tedious algebra, the following second--order differential
equation for $\xi_{mr}$:

\begin{equation}
{\cal A}_2 \frac{d^2 \xi_{mr}}{d r^2} + \frac{{\cal
A}_1}{r} \frac{d \xi_{mr}}{d r} + {\cal A}_0
\frac{\xi_{mr}}{r^2} = \frac{1}{c^2} \frac{d \Psi'_m}{d
r} - {\cal S}_0 \Psi'_m ,
\label{EQ2}
\end{equation}

\noindent with:

\begin{equation}
{\cal A}_2 = 1 + \frac{1}{ \beta} \left( 1 - \frac{c^2}{r^2
\sigma^2} \right),
\label{coefA2}
\end{equation}

\begin{equation}
{\cal A}_1 = \frac{2 c^2}{r^2 \sigma^2 - c^2} \left( -1 -
\frac{\kappa^2}{2 \Omega \sigma} + \frac{ 2 \Omega}{\sigma} +
\frac{r^2 \sigma^2}{c^2} \; c_1 \right) + 1 + d_1 + \frac{ b_1 -1 }
{\beta} \left( 1 - \frac{c^2}{r^2 \sigma^2} \right) ,
\end{equation}

\begin{eqnarray}
{\cal A}_0 = & & \frac{r^2 \sigma^2}{c^2} \left( m^2 -
\frac{\kappa^2}{\sigma^2} \right) - d_1^2 + d_2 + 1 - m^2 + 2 c_1 d_1
+ \frac{2 \Omega}{\sigma} \; d_1 \nonumber \\ & + & \frac{2 r^2
\sigma^2}{r^2 \sigma^2 - c^2} \left( - \frac{\kappa^2}{2 \Omega
\sigma} + \frac{ 2 \Omega}{\sigma} -1 + c_1 \right) \left( 1 - \frac{2
\Omega}{\sigma} + \frac{c^2}{r^2 \sigma^2} d_1 \right) \nonumber \\ &
+ & \frac{1}{\beta} \left\{ b_1 \left[ \frac{2 \Omega}{\sigma} -
\frac{1}{2} - \frac{d_1}{2} \left( 1 + \frac{c^2}{r^2 \sigma^2}
\right) + \frac{c^2}{r^2 \sigma^2- c^2} \left( - \frac{\kappa^2}{2
\Omega \sigma} + \frac{2 \Omega}{\sigma} - 1 + c_1 \right) \right]
\right. \nonumber \\ & & \left.  \; \; \; \; \; \; \; \; \; +
\frac{b_2}{2} - m^2 + \frac{c^2}{r^2 \sigma^2} \left( m^2-1 \right) -
\frac{ b_1^2 c^2}{4 \beta r^2 \sigma^2 } \right\},
\end{eqnarray}

\begin{equation}
{\cal S}_0 = \frac{1}{r^3 \sigma^2} \left[ \frac{2 r^2 \sigma^2 }{r^2
\sigma^2 - c^2} \left( \frac{\kappa^2}{2 \Omega \sigma} - \frac{2
\Omega}{\sigma} + 1 - c_1 \right) - \frac{2 r^2 \Omega \sigma }{c^2} +
\frac{b_1}{2 \beta} \right],
\end{equation}

\noindent where $\beta \equiv c^2 / v_A^2$, with $v_A \equiv
\sqrt{\langle B^2 \rangle / \left( \mu_0 \Sigma \right)}$ being the
Alfv\'en speed, and we have defined the dimensionless quantities:

\begin{equation}
d_1  \equiv  \frac{d \ln \Sigma}{d \ln r} , \; \;
d_2  \equiv  \frac{r^2}{\Sigma} \frac{d^2 \Sigma}{d r^2} , \; \;
c_1  \equiv  \frac{d \ln c}{d \ln r}.
\label{cd}
\end{equation}

The magnetic field comes in through the parameter $\beta$ only.

We see by multiplying all the coefficients of equation~(\ref{EQ2}) by
$\sigma^2$ that there is no singularity in the coefficients of this
equation where $\sigma=0$, i.e. at corotation, in contrast to the
nonmagnetic case (GT79).  When there is no magnetic field, the
particles located at corotation can respond secularly to a forcing
with frequency $\Omega=\Omega_p$ and therefore stay exactly in phase
with the perturbation (they 'surf' the tidal wave).  In the presence
of a toroidal field however, the tension of the field line to which
the particles are attached provides a restoring force which introduces
some inertia in the response of the particles, destroying the
resonance.

From equations~(\ref{Fpertphi}), (\ref{vphim}) and~(\ref{Wm}), we can
write $W'_m$ and $v'_{m \varphi}$ as functions of $\xi_{mr}$.  We give
these expressions as they will be needed below:

\begin{equation}
W'_m = \frac{r^2 \sigma^2 c^2}{r^2 \sigma^2 - c^2} \left[ -
\frac{d \xi_{mr}}{d r} + {\cal C} \frac{\xi_{mr}}{r} +
\frac{\Psi'_m}{r^2 \sigma^2} \right] ,
\label{Wm2}
\end{equation}

\noindent with: 

\begin{equation}
{\cal C} =  - \frac{d}{d \ln r} \left[ \ln \left(
\Sigma r \sigma \right) \right] + \frac{\kappa^2}{2 \Omega \sigma} -
\frac{b_1 c^2}{2 \beta r^2 \sigma^2}, 
\end{equation}

\noindent and:

\begin{equation}
v'_{m \varphi} = \frac{r^2 \sigma^2 c}{r^2 \sigma^2 - c^2} \left[
\frac{c}{r \sigma} \frac{d \xi_{mr}}{d r} + {\cal D}
\frac{\xi_{mr}}{r} - \frac{\Psi'_m}{r \sigma c} \right] ,
\label{vphim2}
\end{equation}

\noindent with: 

\begin{equation}
{\cal D} = \frac{c}{r \sigma} \left\{ \frac{d}{d \ln r}
\left[ \ln \left( \Sigma r \sigma \right) \right] + \frac{b_1}{2
\beta} \right\} - \frac{\kappa^2 r}{2 \Omega c}.
\end{equation}

\subsection{WKB dispersion relation}
\label{sec:wkb}

We assume that the first--order derivative term in
equation~(\ref{EQ2}) can be neglected (this is verified {\it a
posteriori}) and we define $x \equiv r/r_p - 1$, where $r_p$ is the
corotation radius, i.e., for a circular orbit, the planet orbital
radius.  Then the homogeneous equation associated with
equation~(\ref{EQ2}) can be rewritten as:

\begin{equation}
\left[ 1 + \frac{1}{ \beta} \left( 1 - \frac{c^2}{r^2 \sigma^2}
\right) \right] \frac{d^2}{d x^2} \left( \frac{\xi_{mr}}{r_p}
\right) + \frac{{\cal A}_0}{\left(1+x \right)^2} \frac{\xi_{mr}}{r_p} = 0.
\label{EQ2WKB}
\end{equation}

\noindent 
We assume $|x| \ll 1$ and $c \ll r \left| \sigma \right|$, which means
$\left| x \right| \gg H/r$, where $H \sim c / \Omega $ is the disc
semi--thickness, and $\beta \sim 1$.  With $1+x$ being typically on
the order of unity, the WKB approximation can then be used if $\left|
{\cal A}_0 \right| \gg 1$.  We have:

\begin{displaymath}
{\cal A}_0 \simeq 
\frac{r^2 \sigma^2}{c^2} \left( m^2 - \kappa^2 / \sigma^2 \right) ,
\end{displaymath}

\noindent providing $1 - \kappa^2 / \left( m^2 \sigma^2 \right) \gg
c^2/ \left( r^2 \sigma^2 \right)$, i.e. $x^2 \gg H^2/r^2 + 1/m^2$.
When this condition is satisfied then $\left| {\cal A}_0 \right| \gg
1$ and we may look for homogeneous free--wave solutions of the form
$\xi_{mr}(r)=A(r) e^{ikr}$, where $kr \gg 1$ and $A$ is a slowly
varying amplitude.  Since ${\cal A}_1$ is of order $(H/r)^2 x^{-3}$,
the first--order derivative term can be neglected compared to the
second--order derivative term in equation~(\ref{EQ2}) if $x^3 \gg
(H/r)^2 (kr)^{-1}$.  When this is satisfied, the assumption made in
this subsection is validated.  Then equation~(\ref{EQ2WKB}) yields:

\begin{equation}
k = \pm \sqrt{ \frac{m^2 \sigma^2 - \kappa^2}{c^2 + v_A^2}} ,
\label{WKBk}
\end{equation}

\noindent which is the dispersion relation for magnetosonic waves
(e.g., Tagger et al. 1990) propagating outside the Lindblad resonances
(located at $r=r_L$ where $m^2 \sigma^2 - \kappa^2 =0$) with the group
velocity:

\begin{equation}
v_g = - \frac{k \left( c^2 + v_A^2 \right)}{m \sigma} .
\end{equation}

\noindent 
In the absence of magnetic field, these waves reduce to density waves
(GT79).

These solutions describe the response of the disc in the parts where
the perturbing potential is negligible, i.e. outside the Lindblad
resonances and far enough from them.  

We now consider the response of the disc inside the Lindblad
resonances.

\subsection{Magnetic resonances}

\label{sec:magneticres}

The differential equation~(\ref{EQ2}) is singular at the radii $r_M$
where ${\cal A}_2$ vanishes, i.e.  where $1+ \left[ 1 - c^2/ \left(
r^2 \sigma^2 \right) \right] /\beta = 0$.  We call these locations
{\it magnetic resonances}.  There are two such radii, the {\it inner}
magnetic resonance $r_{IMR} < r_p$ and the {\it outer} magnetic
resonance $r_{OMR} > r_p$.  At these locations:

\begin{equation}
m^2 \left( \Omega - \Omega_p \right)^2 = \frac{m^2 c^2 v_A^2}{r^2 
\left( v_A^2 + c^2 \right)} ,
\end{equation}

\noindent i.e. the frequency of the perturbation in a frame rotating
with the fluid matches that of a slow MHD wave propagating along the
field line.  

Using $\Omega = \Omega_K \left[ 1 + {\cal O} \left( H^2/r^2 \right)
\right]$, where $\Omega_K$ is the Keplerian angular velocity, and $H/r
\ll 1$ we get, to first order in $H/r$:

\begin{equation}
\left| r_M - r_p \right| = \frac{2 H}{3 \sqrt{1+ \beta}} ,
\end{equation}

\noindent where $H$ and $\beta$ are evaluated in the vicinity of
$r_p$.  We suppose here that these quantities vary on a scale large
compared to $H$.  As the field becomes weaker ($\beta \rightarrow
\infty$), the magnetic resonances converge toward the corotation
radius.  Note that since the effective Lindblad resonances are all
located beyond $2H/3$ from corotation (see section~\ref{sec:turning}),
they are beyond the magnetic resonances.

We define the new variable $x \equiv \left( r-r_M \right)/ r_M$, where
$r_M$ is either the inner or outer resonance, and we do a local
analysis of equation~(\ref{EQ2}) around $x=0$.  For $\left| x \right|
\ll 1$, we have ${\cal A}_2 \simeq (d{\cal A}_2/dr)_{r=r_M}(r-r_M)$.
Here we assume that $h \equiv c/ \left( r \Omega \right) \sim H/r \ll
1$, the parameters $b_1$, $b_2$, $c_1$, $d_1$ and $d_2$ are at most of
order unity and $\beta \sim 1$.  This amounts to assuming that only
$\sigma$ varies in the expression~(\ref{coefA2}) of ${\cal A}_2$.  For
$\left| x \right| \ll 1$, we then have:

\begin{equation}
{\cal A}_2 \simeq 2 \epsilon \; \frac{\left( 1 + \beta_M
\right)^{3/2}}{\beta_M} \; \left( - \frac{d \ln \Omega}{d \ln r}
\right)_{r_M} \; \frac{x}{h_M} ,
\end{equation}

\noindent where the subscript $M$ indicates that the corresponding
quantities are taken at $r=r_M$, and $\epsilon = -1$ or $+1$ depending
on whether $r_M=r_{IMR}$ or $r_{OMR}$.

We now approximate ${\cal A}_1$, ${\cal A}_0$ and ${\cal S}_0$ by
their values at $r=r_M$.  This requires $\left| x \right| \ll h_M$.  If
this condition is not satisfied, $\Omega-\Omega_p$ cannot be replaced
by its value at $r=r_M$.  To the first non zero order in $h_M$ and
allowing for $m \sim h_M^{-1}$, we can then rewrite
equation~(\ref{EQ2}) under the form:

\begin{equation}
x \frac{\partial^2 \tilde{\xi}_{mr}}{\partial x^2} + \frac{\partial
\tilde{\xi}_{mr}}{\partial x} + {\cal A} \tilde{\xi}_{mr} = {\cal S} ,
\label{EQ2b}
\end{equation}

\noindent where we have defined the dimensionless quantity
$\tilde{\xi}_{mr} \equiv \xi_{mr}/r_M$ and where ${\cal A}$ and ${\cal
S}$ are given by:

\begin{equation}
{\cal A} =  \frac{\epsilon \beta_M}{3 \left( 1 + \beta_M
\right)^{3/2} h_M} \left( \frac{m^2 h_M^2}{1+\beta_M} + 5 + \frac{6}{\beta_M}
\right),
\end{equation}

\begin{equation}
{\cal S} = \frac{1}{3 \left( 1 + \beta_M \right) h_M} \left[ 
\frac{\epsilon \beta_M}{\left( 1 + \beta_M \right)^{1/2}} \left(
\frac{1}{r^2 \Omega^2} \frac{\partial \Psi'_m}{\partial x}
\right)_{x=0} + \frac{3 + \beta_M}{h_M} \left( \frac{\Psi'_m}{r^2
\Omega^2} \right)_{x=0} \; \right].
\end{equation}

\noindent We next introduce the new variable $u=2 \sqrt{{\cal A}x}$,
so that equation~(\ref{EQ2b}) becomes:

\begin{equation}
\frac{\partial^2 \tilde{\xi}_{mr}}{\partial u^2} + \frac{1}{u}
\frac{\partial \tilde{\xi}_{mr}}{\partial u} + \tilde{\xi}_{mr} =
\frac{{\cal S}} {{\cal A}}.
\label{EQ2c}
\end{equation}

Note that $u$ is real outside the magnetic resonances, where ${\cal
A}x>0$, whereas it is imaginary inside the resonances, where ${\cal
A}x<0$.  The solutions of this equation can be written as:

\begin{equation}
\tilde{\xi}_{mr} \left( u \right) = \frac{{\cal S}}{{\cal A}}
+ \frac{\pi}{2} C_\epsilon Y_0(u) + C'_\epsilon J_0(u) ,
\label{solOMR}
\end{equation}

\noindent where $J_0$ and $Y_0$ are the Bessel functions of first and
second kind, respectively, and $C_\epsilon$ and $C'_\epsilon$ are
(complex) constants which {\it a priori} depend on $\beta_M$, $h_M$
and $m$.  The factor $\pi/2$ has been introduced to simplify the
asymptotic expression of $\tilde{\xi}$ below.

When $u \rightarrow 0$ (i.e. $\left| x \right| \ll h_M$), $J_0(u) \sim
1$ and $Y_0(u) \sim (2 / \pi) \ln u$ (Abramowitz \& Stegun 1972).  To
deal with the singularity at $u=0$, we use the Landau prescription,
i.e. we displace the pole slightly from the real axis by replacing
$\Omega - \Omega_p$ with $\Omega - \Omega_p - i \Gamma$, where
$\Gamma$ is a small positive real constant.  This amounts to
multiplying all the perturbed quantities by ${\rm{exp}}\left(\Gamma t
\right)$, and therefore to considering the response of the disc to a
slowly increasing perturbation.  This procedure can also be thought of
as introducing a viscous force, or as shifting the resonances
progressively with time.  Note that such a shift occurs when the
planet drifts through the disc, since the material in the resonances
then changes with time.  Shifting $\Omega - \Omega_p$ below the real
axis is equivalent to shifting $x$ above this axis in
equation~(\ref{EQ2b}), since only $\Omega-\Omega_p$ was allowed to
vary in the expression of ${\cal A}_2$.  Therefore, we replace $x$
with $x+ i \gamma$, where $\gamma$ is a small positive real constant.
For $\left| x \right| \ll h_M$, the solutions of equation~(\ref{EQ2c})
can then be written as:

\begin{equation}
\tilde{\xi}_{mr} \simeq C_\epsilon \ln \left[ 4 {\cal A} \left( x + i
\gamma \right) \right],
\label{sol0}
\end{equation}

\noindent so that, near $x=0$: 

\begin{equation}
\tilde{\xi}_{mr} \simeq C_\epsilon \left( \ln \left| 4 {\cal A} \gamma
\right| + i \arctan \frac{\gamma}{x} \right).
\label{sol00}
\end{equation}

\noindent Note that there is a phase shift of $\pi$ on passing through
$x=0$.

Since ${\cal S}/ {\cal A}$ has opposite sign at $r=r_{IMR}$ and
$r=r_{OMR}$ ($\partial \Psi'_m / \partial x$ changes sign on passing
through corotation whereas $ \Psi'_m $ does not) and is real, the real
part of the solutions of equation~(\ref{EQ2c}) should have opposite
sign at $r=r_{IMR}$ and $r=r_{OMR}$ whereas their imaginary parts
should have the same sign.  Therefore, from equation~(\ref{sol00}) we
get $C_{+1} = -C_{-1}^{\ast}$, where the asterisk denotes the complex
conjugate.

Equations~(\ref{Wm2}) and~(\ref{vphim2}) can now be used to evaluate
$v'_{m \varphi}$ and $W'_m$ for $\left| x \right| \ll h_M$:

\begin{equation}
\frac{W'_m}{r^2_M \Omega^2_M} \simeq   C_\epsilon 
\frac{h^2_M}{\beta_M \left( x + i \gamma \right)} ,
\label{Wmm}
\end{equation}

\begin{equation}
\frac{v'_{m \varphi}}{r_M \Omega_M} \simeq  \epsilon
C_\epsilon \frac{\left(1 + \beta_M \right)^{1/2} h_M}{\beta_M \left( x
+ i \gamma \right)} ,
\label{vMphi}
\end{equation}

\noindent where $\Omega_M$ is $\Omega$ taken at $r=r_M$.  Finally,
$v'_{mr} = i m \sigma \xi_{mr}$ yields, for $\left| x \right| \ll
h_M$:

\begin{equation}
\frac{v'_{m r}}{r_M \Omega_M} \simeq - \epsilon i
C_\epsilon \frac{m h_M}{\left( 1 + \beta_M \right)^{1/2}} \; \ln
\left[ 4 {\cal A} \left( x + i \gamma \right) \right].
\label{vMr}
\end{equation}

We note that, in a nonmagnetized disc, the azimuthal perturbed
velocity varies as $\ln x$ with $x= \left| r - r_p \right| / r_p
\rightarrow 0$ near corotation (GT79), whereas here it varies as $1/x$
near the magnetic resonances.

\subsection{Turning points}

\label{sec:turning}

At the locations where the solutions of equation~(\ref{EQ2}) change
from wave--like to non wave--like (turning points), the disc response
varies on a large scale, and therefore it couples well to the
perturbing potential.  This results in a large contribution to the
tidal torque (see GT79 and \S~\ref{sec:torque}).  It is therefore
important to calculate the location of these turning points.

The WKB analysis above (see \S~\ref{sec:wkb}) indicates that the
differential equation~(\ref{EQ2}) has turning points at the locations
of the so--called {\it nominal} Lindblad resonances $r=r_L$.  Waves
propagate outside these resonances, which are located on both sides of
the corotation radius.  In the nonmagnetic case, it is known that
the turning points do no co\"{\i}ncide with the nominal Lindblad
resonances for values of $m$ in excess of $r/H$ (Goldreich \& Tremaine
1980, Artymowicz 1993).  If they did co\"{\i}ncide, the turning points
(also called {\it effective} Lindblad resonances) would converge
toward $r_p$ as $m$ increases.  Instead, they converge toward a radius
which is located a finite distance (equal to $2H/3$) away from $r_p$.
Since the potential is more and more localized around $r_p$ at large
$m$, coupling between the perturbation and the disc response is lost,
and the contribution to the total torque becomes negligible.  This is
the torque cutoff first discussed by Goldreich \& Tremaine (1980).

The WKB analysis above indicates that for values of $m$ up to about
$r/H$, the location of the turning points does not depend on whether
there is a magnetic field or not.  In both cases they co\"{\i}ncide
with the nominal Lindblad resonances.  However, for larger values of
$m$, there are differences between a magnetized and a nonmagnetized
disc, as the analysis below shows.  We now calculate the location of
the turning points.

\subsubsection{Elimination of the first--order derivative term in 
equation~(\ref{EQ2}):}

There are no singularities in the coefficients of equation~(\ref{EQ2})
at $r=r_L$.  This is similar to the nonmagnetic case, where only the
second--order differential equation for $W'_{m}$ has (removable)
singularities there (Narayan, Goldreich \& Goodman 1987).  All the
physical quantities are regular at these locations.  To identify the
turning points, we eliminate the first--order derivative term in this
equation.  We define $x \equiv (r-r_L)/r_L$, $r_L$ being either the
inner or outer nominal Lindblad resonance, and introduce the new
dependent variable:

\begin{equation}
y = \xi_{mr} \; {\rm exp} {\left( \frac{1}{2} \int \frac{{\cal
A}_1}{{\cal A}_2} \; dx \right)} ,
\end{equation}

\noindent so that the homogeneous equation associated with
equation~(\ref{EQ2}) can be written under the exact form:

\begin{equation}
\frac{d^2 y}{d x^2} + {\cal K} y =0,
\label{EQ2turn}
\end{equation}

\noindent where:

\begin{equation}
{\cal K} = \frac{{\cal A}_0}{{\cal A}_2} - \frac{1}{4} \left(
\frac{{\cal A}_1}{{\cal A}_2} \right)^2 - \frac{1}{2} \frac{d}{dx}
\left( \frac{{\cal A}_1}{{\cal A}_2} \right) .
\end{equation}

We now expand ${\cal K}$ around $x=0$.  At $r=r_L$, $\kappa^2 = m^2
\sigma^2$.  We neglect here the departure from Keplerian rotation
(which is second order in $H/r$) so that $\kappa = \Omega$.  Then
$\sigma = \epsilon \Omega(r_L) X /m$, with $X=1 + 3 \epsilon m x /2$
and $\epsilon=+1$ or $-1$ depending on whether we consider the outer
or inner resonance, respectively.  We assume $\beta$ and $h \equiv
c/(r \Omega) \sim H/r$ constant.  Note that for values of $m \sim 1/h$
and for $x \sim h$, we cannot approximate $\sigma$ by its value at
$r_L$.  After some algebra, we find, to the first non zero order:

\begin{eqnarray}
\frac{m^2}{{\cal A}^2_2} \; {\cal K} = & &  \frac{X^2 -1}{m^2 h^2} -1 -
\frac{6}{X^2 -m^2 h^2} - \frac{27}{4} \frac{m^2 h^2}{\left( X^2 -
m^2 h^2 \right)^2} \nonumber \\
& + & \frac{1}{\beta} \left\{  \left( 1- \frac{m^2
h^2}{X^2} \right) \left[ \frac{X^2 -1}{m^2 h^2} - 2 - \frac{6}{X^2
-m^2 h^2} - \frac{1}{\beta} \left( 1 - \frac{m^2 h^2}{X^2} \right)
\right] \right. \nonumber \\
& & \left. \; \; \; \; \; \; - \frac{9 m^2 h^2}
{2 X^2 \left( X^2 - m^2 h^2 \right)}
\left( \frac{3}{2} + \frac{m^2 h^2}{2 X^2} \right) \right\} ,
\label{Kcoef}
\end{eqnarray} 

\noindent where we have assumed $m \gg 1$, since these are the values
of interest for the tidal torque (see GT79).  This expression is valid
even for $m$ larger than or very large compared to $1/h$, and
whatever the values of $b_1, b_2, c_1, d_1$ and $d_2$ compatible with
$\beta$ and $h$ being constant and provided they are not large
compared to unity.  Note that the expression above can also be seen as
an expansion around the corotation radius, which corresponds to $X=0$.

\subsubsection{Calculation of the turning points:}

\begin{figure}
\centerline{
\epsfig{file=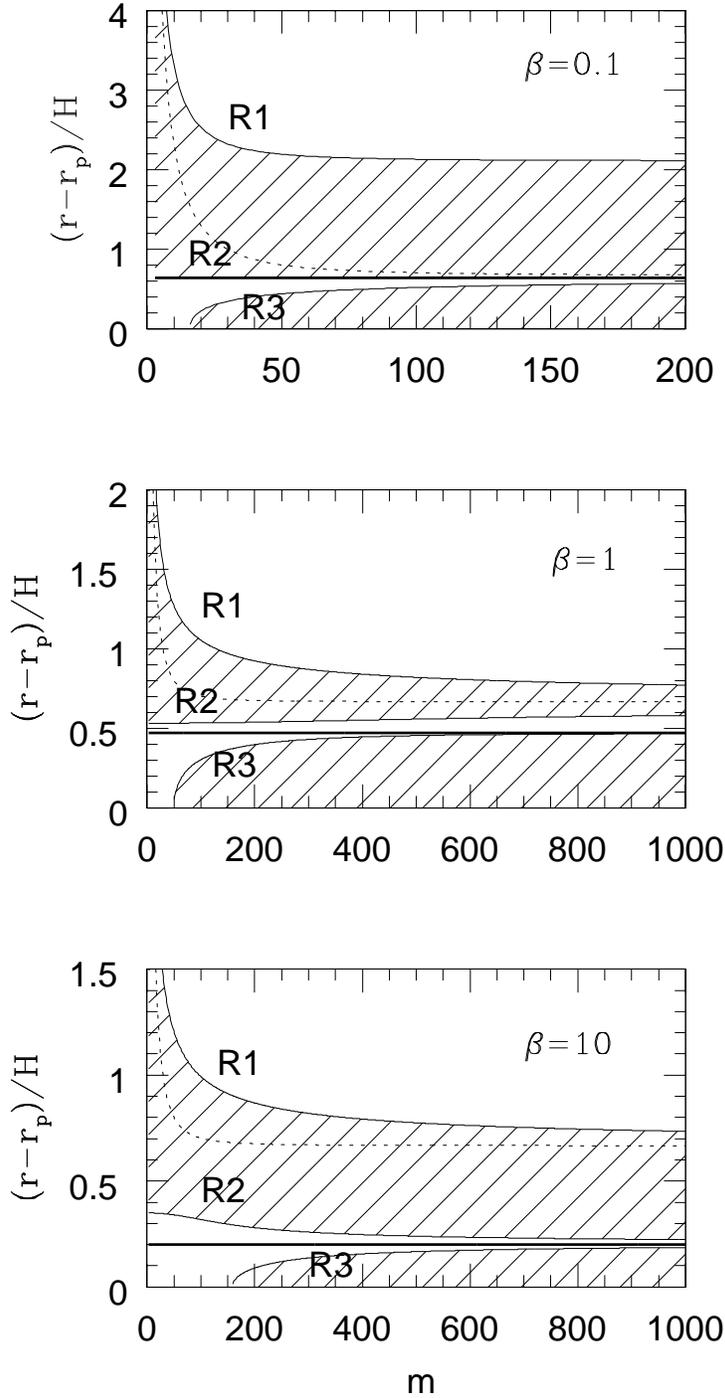} }
\caption[]{Distance of the outer turning points ({\it thin solid
lines}) and magnetic resonance ({\it thick bold lines}) to corotation
in units of $H \equiv c/\Omega$ versus $m$ for $\beta=0.1$ ({\it upper
plot}), 1 ({\it middle plot}) and 10 ({\it lower plot}) and for $H/r
=0.03$.  The turning points are labelled R1, R2 and R3.  The shaded
areas indicate the regions where the waves are evanescent.  The
location of the effective outer Lindblad resonance in a nonmagnetized
disc is also represented ({\it dashed lines}). The inner turning
points and resonance are obtained by reflection with respect to the
$x$--axis.}
\label{fig1}
\end{figure}

The turning points are the radii where ${\cal K}=0$.  We solve this
equation numerically and display the results in figure~\ref{fig1},
where we plot $(r-r_p)/H$ (where $H \equiv c/\Omega$) as a function of
$m$ for the turning points located beyond corotation, and for
$\beta=0.1$, 1 and 10 and $H/r=0.03$.  The outer magnetic resonance is
also represented.  To each of these outer turning points and resonance
corresponds an inner turning point or resonance located at the same
distance from corotation to first order in $H/r$ or $1/m$.  Similar
plots are obtained for other values of $H/r$.

Artymowicz (1993) has calculated that, in a nonmagnetized disc, 
the location of the turning points is given by (his eq.~[39]):

\begin{displaymath}
\frac{r - r_p}{H} = \frac{2}{3 m h_L} \sqrt{1 + m^2 h_L^2} ,
\end{displaymath}

\noindent so that, for $m \rightarrow \infty$, the distance to corotation is 
$2H/3$.

In a magnetized, we identify up to three turning points, that we label
R1, R2 and R3:
\begin{enumerate}
\item The outermost turning point R1 co\"{\i}ncides with the nominal
Lindblad resonance for small values of $m$.  When $m \rightarrow
\infty$, its distance to corotation is $2H/3$ for $\beta \ge 1$ (like
in the nonmagnetic case) and $2H/(3 \sqrt{\beta})$ for $\beta \le 1$,
i.e. $2H'/3$ with $H' \equiv v_A/\Omega$.
\item The intermediate turning point R2 is located at a distance from
corotation equal to $2H/(3 \sqrt{\beta})$ for $\beta \ge 1$ and $2H/3$
for $\beta \le 1$.
\item The innermost turning point R3 exists only above some critical
value of $m$, $m_{\rm crit}$, which depends on $\beta$ and $H/r$.
Interpolation of the numerical results gives $m_{\rm crit} \simeq 1.5
\sqrt{\beta} \times r/H $.  When $m \rightarrow \infty$, the distance
of this turning point to corotation is $2H/( 3 \sqrt{\beta+1})$,
i.e. it co\"{\i}ncides with the magnetic resonance.  For either $\beta
\ll 1$ or $\beta \gg 1$, R2 and R3 merge at large $m$.
\end{enumerate}

In figure~\ref{fig1}, the shaded areas indicate where the waves are
evanescent (i.e. ${\cal K}<0$).  In contrast to the nonmagnetic case,
there is a region inside the outermost turning points where waves
propagate.  This region moves toward the corotation radius and
decreases in size as $\beta$ increases.  It disappears altogether for
infinite $\beta$.  Note that the magnetic resonances are contained
within this region, so that the singular slow modes excited at the
magnetic resonances can propagate.

Since waves do not propagate around corotation for $m>m_{\rm crit}$,
we expect a torque cutoff at large $m$ like in the nonmagnetic case.

We can check whether waves propagate or not around corotation by
taking the limit $X \rightarrow 0$ (i.e. $r \rightarrow r_p$) in the
expression~(\ref{Kcoef}) above.  In the nonmagnetic case ($\beta
\rightarrow \infty$):
$$\frac{m^2} { {\cal A}_2^2} \; {\cal K} \rightarrow -1 - \frac{7}{4
m^2 h_L^2}.$$ Since this is negative, the waves are evanescent around
corotation in that case.  When $\beta$ is finite: $$\frac{m^2 }{ {\cal
A}_2^2} \; {\cal K} \rightarrow \frac{(m^2 h_L^2)^2}{X^4} \left( -
\frac{1}{\beta} + \frac{9}{4 m^2 h_L2} \right).$$ This is positive
when $m<m_{\rm crit}$ with $m_{\rm crit} = 3 \sqrt{\beta} /(2 h_L)$,
so that waves propagate around corotation for these values of $m$.
Note that the value of $m_{\rm crit}$ found by interpolation of the
numerical results agrees perfectly with the analytical expression.

\section{Tidal torque}
\label{sec:torque}

\subsection{Expression for the torque}

In the linear regime, the net torque exerted by the planet on the
disc between the radii $r_1$ and $r_2$ is:

\begin{displaymath}
{\bf T} \left( r_1, r_2 \right) = \sum_{m=0}^{\infty} {\bf T}_m \left(
r_1, r_2 \right),
\end{displaymath}

\noindent with:

\begin{equation}
{\bf T}_m \left( r_1, r_2 \right) = - \int^{r_2}_{r_1} \int^{2 \pi}_0
{\rm Re} \left[ \Sigma + \Sigma'_m e^{i m \left( \varphi - \Omega_p t
\right)} \right] {\bf r} \mbox{\boldmath $\times$} {\rm Re} \left[
\mbox{\boldmath $\nabla$} \left( \Psi'_m e^{i m \left( \varphi -
\Omega_p t \right)} \right) \right] \; r d\varphi dr .
\end{equation}

\noindent Because of the $\varphi$--periodicity, the first--order term
is zero and the torque has only a $z$--component which can be written
as:

\begin{equation}
T_m \left( r_1, r_2 \right) = 2 \pi \int^{r_2}_{r_1} \frac{d T_m}{dA}
\; r dr 
\label{torquer1r2}
\end{equation}

\noindent where, following KP93, we have defined the ``torque
density'':

\begin{equation}
\frac{d T_m}{dA} \equiv \frac{m}{2}  \; {\rm Im} \left( \Sigma'^{\ast}_m
\Psi'_m \right) .
\label{torquedens}
\end{equation}

\noindent The asterisk denotes the complex conjugate.

In a nonmagnetized disc, another useful quantity is the angular
momentum flux:

\begin{equation}
F_m (r) = \Sigma r^2 \int_0^{2 \pi} {\rm Re} \left[ v'_{m \varphi}
e^{i m \left( \varphi - \Omega_p t \right)} \right] \; {\rm Re} \left[
v'_{m r} e^{i m \left( \varphi - \Omega_p t \right)} \right] \; d
\varphi ,
\end{equation}

\noindent which can also be written under the form:

\begin{equation}
F_m(r) = \pi \Sigma r^2 {\rm Re} \left( v'_{m r} v'^{\ast}_{m \varphi}
\right) .
\end{equation}

In a nonmagnetized disc, inside the Lindblad resonances and away from
corotation, the response of the disc is non wave--like and in phase
with the perturbation, so that $\Sigma'^{\ast}_m \Psi'_m$ is real.
Outside the Lindblad resonances, the wavelength of the perturbation is
small compared to the scale on which the perturbing potential varies
so that $r \Sigma'^{\ast}_m \Psi'_m$ integrated over radius is small.
Therefore, most of the contribution to the torque comes from the
resonances (Lynden--Bell \& Kalnajs 1972, GT79).  There are two
contributions to the total torque.  These are the Lindblad torque,
which is equal to the angular momentum advected through the disc, and
the corotation torque, which accumulates at the corotation resonance
(GT79) and is associated with a discontinuity of the angular momentum
flux there.

In the presence of a magnetic field, it is not clear it is possible to
calculate the torque separately at the different turning points,
because the scale on which the disc response varies there may become
larger than the distance between the turning points.  In addition, the
torque exerted around corotation for the values of $m$ for which
waves propagate there may be significant, as the scale on which the
perturbing potential varies and that on which the disc response varies
may be similar.  The total torque will then be calculated numerically.

However, an asymptotic form of the torque density at the magnetic
resonances can be obtained, and this is done in the next subsection.

\subsection{Torque at the magnetic resonances}

\label{sec:torquereson}

The torque density in the vicinity of a magnetic resonance is, from
equation~(\ref{torquedens}) and with $\Sigma'_m= \Sigma W'_m / c^2$
given by equation~(\ref{Wmm}):

\begin{eqnarray}
\frac{1}{\left( \Sigma r^2 \Omega^2 \right)_{r_M}} \frac{dT^M_m}{dA} &
= & \left( \frac{ \Psi'_m}{r^2 \Omega^2 } \right)_{r_M} \frac{m}{2
\beta_M} \; {\rm Im} \left( \frac{C^{\ast}_\epsilon}{x - i \gamma}
\right) , \label{torquedensM} \\ 
& = & \left( \frac{ \Psi'_m}{r^2 \Omega^2 } \right)_{r_M} \frac{m}{2
\beta_M} \; \frac{x {\rm Im} \left(
C^{\ast}_\epsilon \right) + \gamma {\rm Re} \left(
C^{\ast}_\epsilon \right) }{x^2 + \gamma^2}  . \label{torquedensM2}
\end{eqnarray}

To get the torque exerted over a magnetic resonance, we integrate the
torque density over the width $\Delta r$ of the resonance, i.e. from
$r_M- \Delta r/2$ to $r_M + \Delta r/2$.  Note that $\Delta x \equiv
\Delta r / r_M \sim \gamma$.  Equation~(\ref{Wmm}) has been used to
derive the above expression for the torque density.  Therefore, only
if $\Delta x$, and hence $\gamma$, is small compared to $h_M$ can this
expression be used to calculate the torque over the resonance.  When
this condition is satisfied, the torque can be written as:


\begin{equation}
\frac{T^M_m}{\left( \Sigma r^4 \Omega^2 \right)_{r_M}} \sim {\rm Re}
\left( C_\epsilon \right) \; \frac{2 \pi m }{\beta_M} \left( \frac{
\Psi'_m}{r^2 \Omega^2 } \right)_{r_M} \arctan \frac{ \Delta x}{2
\gamma}.
\label{torqueres}
\end{equation}

\noindent with $\arctan \left[ \Delta x /(2 \gamma) \right]\sim \pi/2$
if $\Delta x $ is a few times $\gamma$.

Here we have assumed that the resonance were symmetrical around $r_M$,
so that the term proportional to ${\rm Im} \left( C_\epsilon^\ast
\right)$ in equation~(\ref{torquedensM2}) gives no contribution.
Since the integral over the term proportional to ${\rm Re} \left(
C_\epsilon^\ast \right)$ converges (does not depend on the width of
the resonance), the torque we obtain can be considered as a
point--like contribution from the resonance, similar to the corotation
torque in the nonmagnetic case.  However, the resonance may not be
symmetrical, in which case the term proportional to ${\rm Im} \left(
C_\epsilon^\ast \right)$ gives a global contribution, which does
depend on the extent of the resonance.  We will see below that the
torque around the magnetic resonances is indeed dominated by this
global contribution, and not by the point--like torque.

Since ${\rm Re} \left( C_{+1} \right) = - {\rm Re} \left( C_{-1}
\right)$, the torque derived above has opposite sign at the inner and
outer magnetic resonances.

To know the amplitude and the sign of $T^M_m$ we need to calculate
${\rm Im} \left( C_\epsilon \right)$.  This cannot be done from the
local analysis presented in section~\ref{sec:magneticres}, as only the
solutions of the homogeneous equation determine the behaviour of the
disc response at the magnetic resonances.  The solutions obtained
around the magnetic resonances are valid only for $|r-r_M| \ll H$ and
therefore cannot be matched to the WKB solutions calculated in
section~\ref{sec:wkb}.  The constants $C_\epsilon$ will then be
calculated by matching the numerical and analytical solutions, and the
asymptotic form of the torque density given by
equation~(\ref{torquedensM}) will be used to obtain a precise value of
the torque at the magnetic resonances.

\section{Numerical calculations}
\label{sec:numerics}

In order to calculate the tidal torque exerted by the planet on the
disc, we have solved the second--order linearized differential
equation~(\ref{EQ2}) numerically.

\subsection{Numerical scheme}

The numerical scheme we have used is similar to that presented by
KP93.  

The coefficients $\Psi'_m$ of the Fourier decomposition of the
potential (see eq.~[\ref{Psifourier}]) are calculated using the
recurrence formulae between the generalized Laplace coefficients (see
the appendix of KP93, which unfortunately contains several typos).

Equation~(\ref{EQ2}) is then integrated using the fifth--order
Runge--Kutta procedure with adaptative stepsize control given by
Press~et al.~(1992) from the disc inner radius $r_{in}$ to its outer
radius $r_{out}$.  Note that KP93 started the integration at the
corotation resonance and shot toward the boundaries.  This procedure
would not be appropriate here as in the presence of a magnetic field
there is a resonance on each side of corotation.  For a given value of
$m$, we first compute two independent solutions $\xi_{mr,1}$ and
$\xi_{mr,2}$ of the homogeneous equation, and one particular solution
$\xi_{mr,p}$ of the inhomogeneous equation starting with random
initial values.  The general solution is then given by $\xi_{mr}=
\xi_{mr,p} + c_1 \xi_{mr,1} + c_2 \xi_{mr,2}$, where $c_1$ and $c_2$
are constants which depend on the boundary conditions.

To calculate $c_1$ and $c_2$, we use the WKB solution of the equation,
which should be valid at $r_{in}$ and $r_{out}$ if the boundaries
are far enough from corotation.  Then, at these locations, $d
\xi_{mr}/dr= i k \xi_{mr}$, where $k$ is given by
equation~(\ref{WKBk}). Since the physical solutions are outgoing
waves, we take $k>0$.  The group velocity is then
positive (negative) outside (inside) corotation.  These two boundary
conditions yield two algebraic equations for $c_1$ and $c_2$ (see
eq.~[24] of KP93) which can be solved to find these coefficients.

To get better numerical accuracy, we repeat the procedure taking for
the starting value of the particular solution $\xi_{mr,p}$ the value
of $\xi_{mr}$ calculated at $r=r_{in}$ with the above coefficients
$c_1$ and $c_2$.  This leads to $\xi_{mr,p}$ converging toward the
full solution $\xi_{mr}$.  When recalculating $c_1$ and $c_2$ after
one iteration we indeed verify that these two constants are very close
to zero.  This procedure prevents large cancellation between
$\xi_{mr,p}$ and the general solutions of the homogeneous equation.

Note that KP93 found the WKB approximation not to be accurate enough
for the boundary conditions, and they accordingly took into account
the amplitude as well as the phase variation of the WKB solution.  We
found that we could reproduce their results without having to go
beyond the WKB approximation.

Once $\xi_{mr}$ is known, $W'_m$ and $v'_{m \varphi}$ can be
calculated from equations~(\ref{Wm2}) and~(\ref{vphim2}), $v'_{mr}=im
\sigma \xi_{mr}$, and the torque density $dT_m/dA$ is given by
equation~(\ref{torquedens}).  To calculate the total torque $T_m$
exerted by the planet on the disc, we integrate $dT_m/dA$ over the
disc surface (see eq.~[\ref{torquer1r2}]).  

\subsection{Numerical results}

The planet potential softening length is set to $r_0=10^{-4} r_p$.
Note that in the presence of a magnetic field the results are
insensitive to the value of $r_0$ provided it is small enough, since
there is no singularity of the homogeneous equation at $r=r_p$, where
the potential is singular.  The Landau parameter is set to
$\gamma=10^{-6}$.  As expected, the total torque does not depend on
the value of $\gamma$ provided it is small enough.

For $m \le 5$, we integrate the equation from $r_{in}=0.2 r_p$ to
$r_{out}=5 r_p$, whereas, for $m \ge 6$, we limit the range of
integration from $r_{in}=0.5 r_p$ to $r_{out}=1.5 r_p$.

We assume that $\Sigma$, $c$ and $r^2 \langle B^2 \rangle$ vary like
power laws of $r$.  According to equations~(\ref{bs}) and~(\ref{cd}),
we then have $\Sigma \propto r^{d_1}$, $c \propto r^{c_1}$ and $r^2
\langle B^2 \rangle \propto r^{b_1}$. Then $d_2=d_1(d_1-1)$ and
$b_2=(b_1-2)(b_1-1)$.  Note that $H/r \sim c/ \left( r \Omega \right)
\propto r^{c_1+0.5}$ and $\beta \equiv c^2/v_A^2 \propto
r^{2c_1+d_1-b_1+2}$.  We comment that the values of the parameters we
chose below may require a departure from Keplerian rotation that is
second order in $H/r$ to ensure hydrostatic equilibrium.  This is here
neglected, as in Ward~(1986, 1997).

To simplify the discussion, we define the dimensionless quantities
$\tilde{W}'_m = W'_m/\left( r_p^2 \Omega_p^2 \right)$, $\tilde{T}_m =
T_m /\left( \Sigma_p r_p^4 \Omega_p^2 \right)$, $\tilde{T}_{m,D} = 2
\pi (T_m/dA) /\left( \Sigma_p r_p^2 \Omega_p^2 \right)$ and
$\tilde{F}_m = F_m /\left( \Sigma_p r_p^4 \Omega_p^2 \right)$, where
the subscript $p$ indicates that the quantity is taken at $r=r_p$.
The numerical results presented in this section correspond to
$M_p/M_\ast=1$.  Since $\Psi'_m / ( r_p^2 \Omega_p^2 ) \propto
M_p/M_\ast$ (see section~\ref{sec:potential}), then $\tilde{W}'_m
\propto M_p/M_\ast$ and $\tilde{T}_m \propto (M_p/M_\ast)^2$.

\subsubsection{Case $B=0$}

\begin{figure}
\centerline{
\epsfig{file=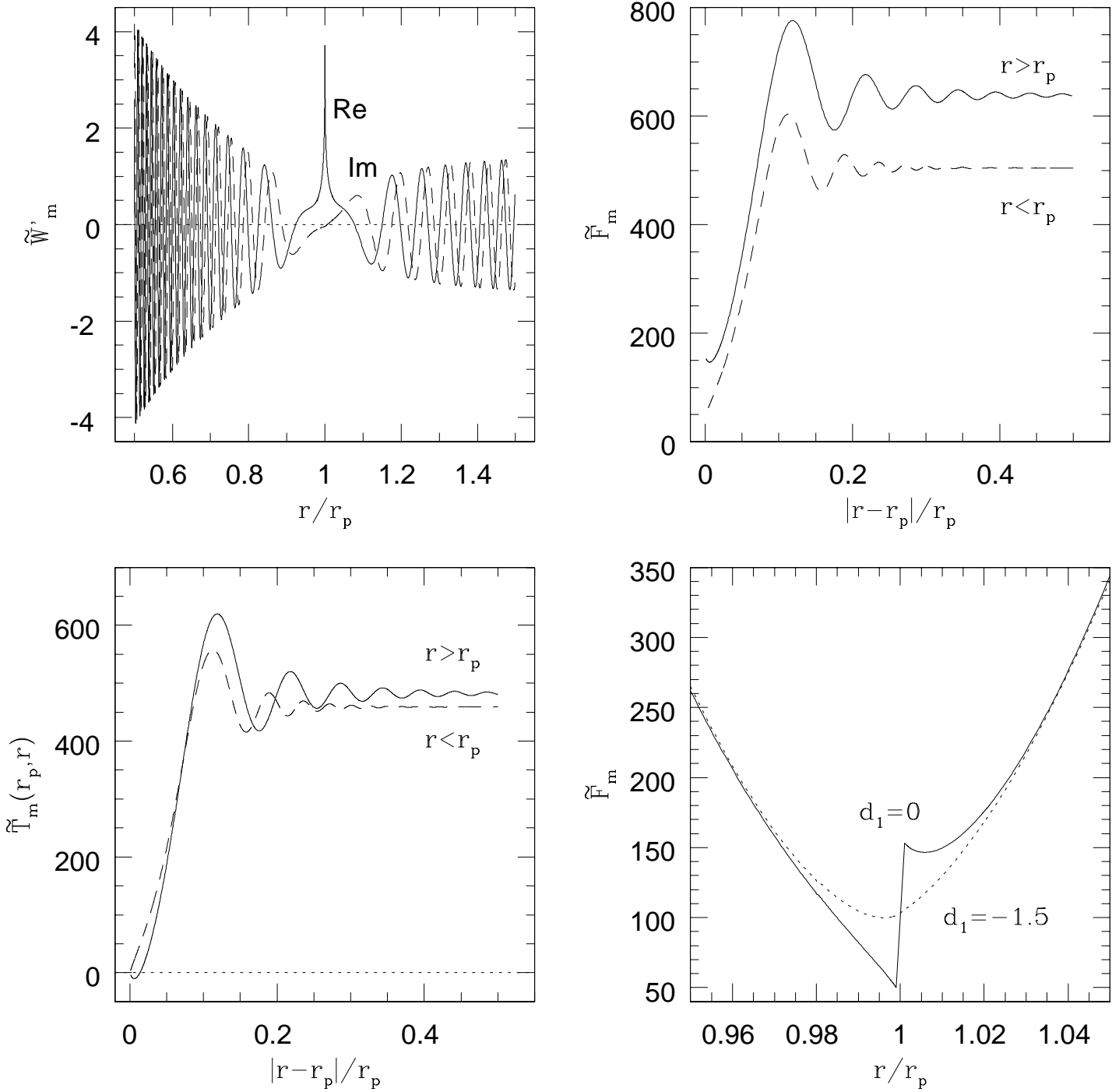} }
\caption[]{Results for $B=0$, $d_1 \equiv d \ln \Sigma/ d \ln r = 0$,
$c/ \left( r_p \Omega_p \right) = 0.03$ ($c_1=0$) and $m=10$.  {\em
Upper left panel:} Real part ({\em solid line}) and imaginary part
({\em dashed line}) of $\tilde{W}'_m$ versus $r/r_p$. {\em Lower left
panel:} Tidal torque $\tilde{T}_m(r_p,r)$ exerted by the planet on the
disc between the radii $r_p$ and $r$ versus $|r-r_p|/r_p$.  The two
curves correspond to $r>r_p$ ({\em solid line}) and $r<r_p$ ({\em
dashed line}). {\em Upper right panel:} Angular momentum flux
$\tilde{F}_m$ versus $|r-r_p|/r_p$.  The two curves correspond to
$r>r_p$ ({\em solid line}) and $r<r_p$ ({\em dashed line}). {\em Lower
right panel:} $\tilde{F}_m$ versus $r/r_p$ around corotation for $d_1=
0$ ({\em solid line}) and -1.5 ({\em dotted line}).}
\label{fig2}
\end{figure}

By setting $B=0$, we have checked that we recover the KP93 results.
In that case, the corotation torque is non zero if $d_1 \ne -1.5$ and
corresponds to a discontinuity of the angular momentum flux at
$r=r_p$.  In figure~\ref{fig2}, we plot $\tilde{W}'_m$, the angular
momentum flux $\tilde{F}_m(r)$ at radius $r$ and the torque
$\tilde{T}_m(r_p,r)$ exerted by the planet on the disc between the
radii $r_p$ and $r$ for $B=0$, $d_1=0$, $c/ \left( r_p \Omega_p
\right) = 0.03$ ($c_1=0$) and $m=10$.  For comparison, we also plot
the angular momentum flux corresponding to $d_1=-1.5$ in the
corotation region.  The parameters used here are the same as those
used by KP93 in their figures~2, 5 and 6, so that a direct comparison
can be made.  For $d_1=0$, the dimensionless corotation torque is
about 102, whereas the dimensionless total torque
$\tilde{T}_m(r_{in},r_{out})$ is about 22, in very good agreement with
the results displayed by KP93.  

\subsubsection{Case $B \ne 0$}

\label{sec:caseBne0}

\begin{figure}
\centerline{ \epsfig{file=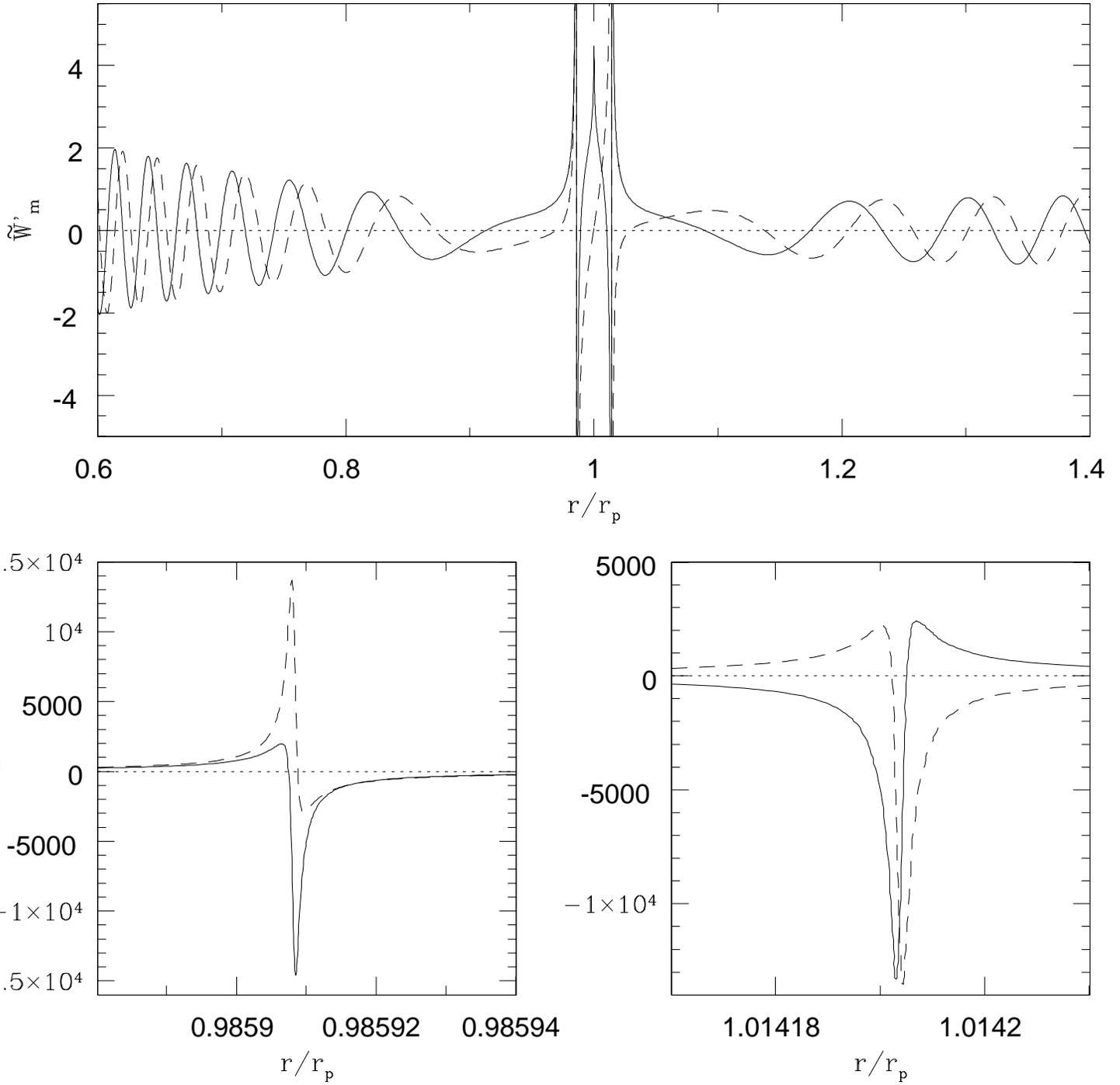} }
\caption[]{Results for $B \ne 0$:  $\tilde{W}'_m$ versus $r/r_p$ for
$d_1=0$, $c/ \left( r_p \Omega_p \right) = 0.03$ ($c_1=0$), $b_1=2$, $
\beta = {\rm const} = 1$ and $m=10$.  Both the real part ({\em solid
line}) and imaginary part ({\em dashed line}) are plotted.  The two
lower panels show a zoom on the magnetic resonances ({\em left panel},
inner resonance; {\em right panel}, outer resonance).  }
\label{fig3}
\end{figure}

\begin{figure}
\centerline{ \epsfig{file=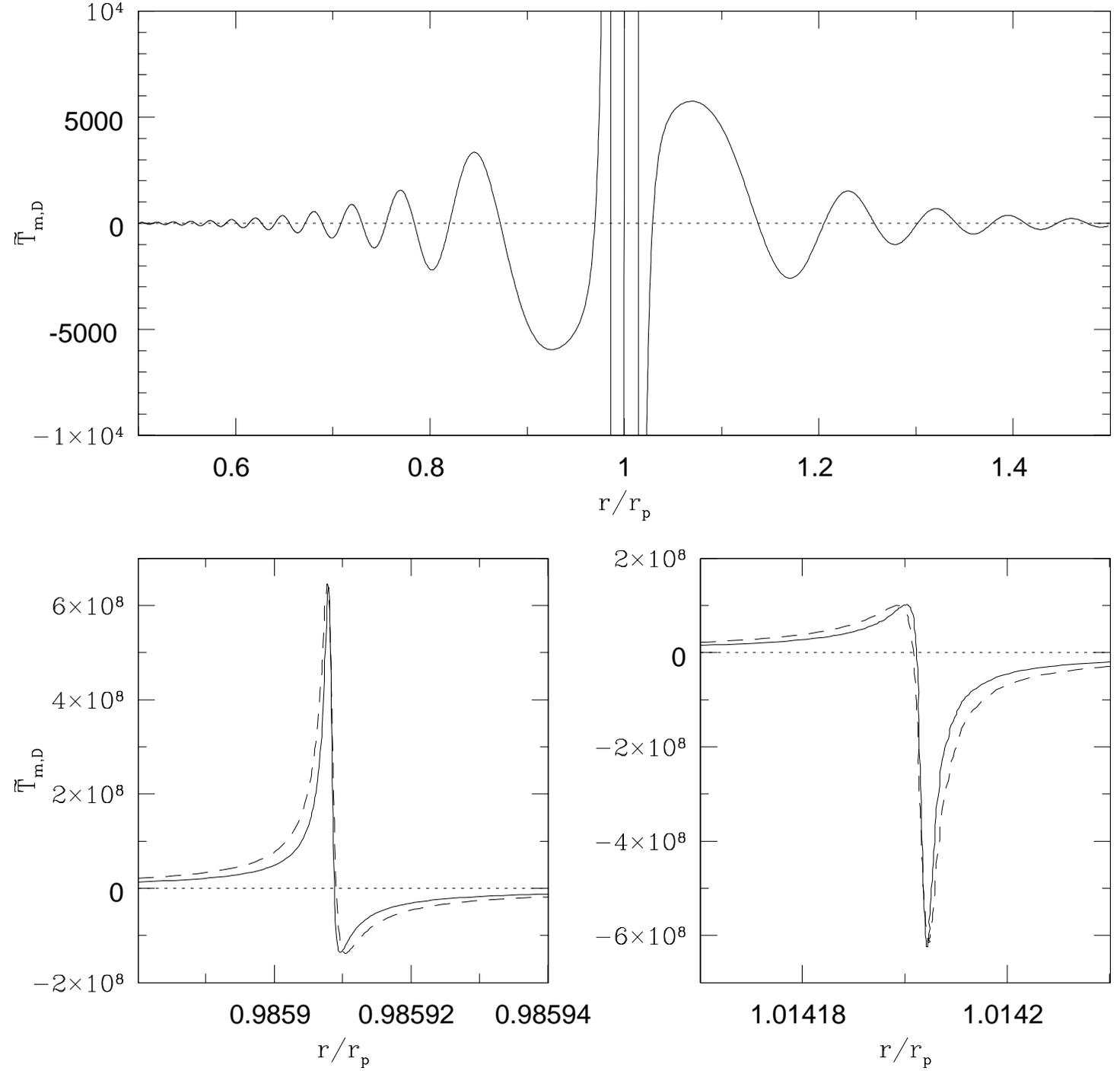} }
\caption[]{Torque density $\tilde{T}_{m,D}$ versus $r/r_p$ for the
same parameters as in fig.~\ref{fig3}.  The two lower panels show a
zoom on the magnetic resonances ({\em left panel}, inner resonance;
{\em right panel}, outer resonance).  The dashed lines represent the
fit given by equation~(\ref{torquedensM}), where the constants
$C_\epsilon$ are calculated by matching $W'_m$ obtained numerically
with the analytical expression~(\ref{Wmm}). The point--like torque
exerted at the inner magnetic resonance is positive whereas that
exerted at the outer magnetic resonance is negative. }
\label{fig4}
\end{figure}

\begin{figure}
\centerline{ \epsfig{file=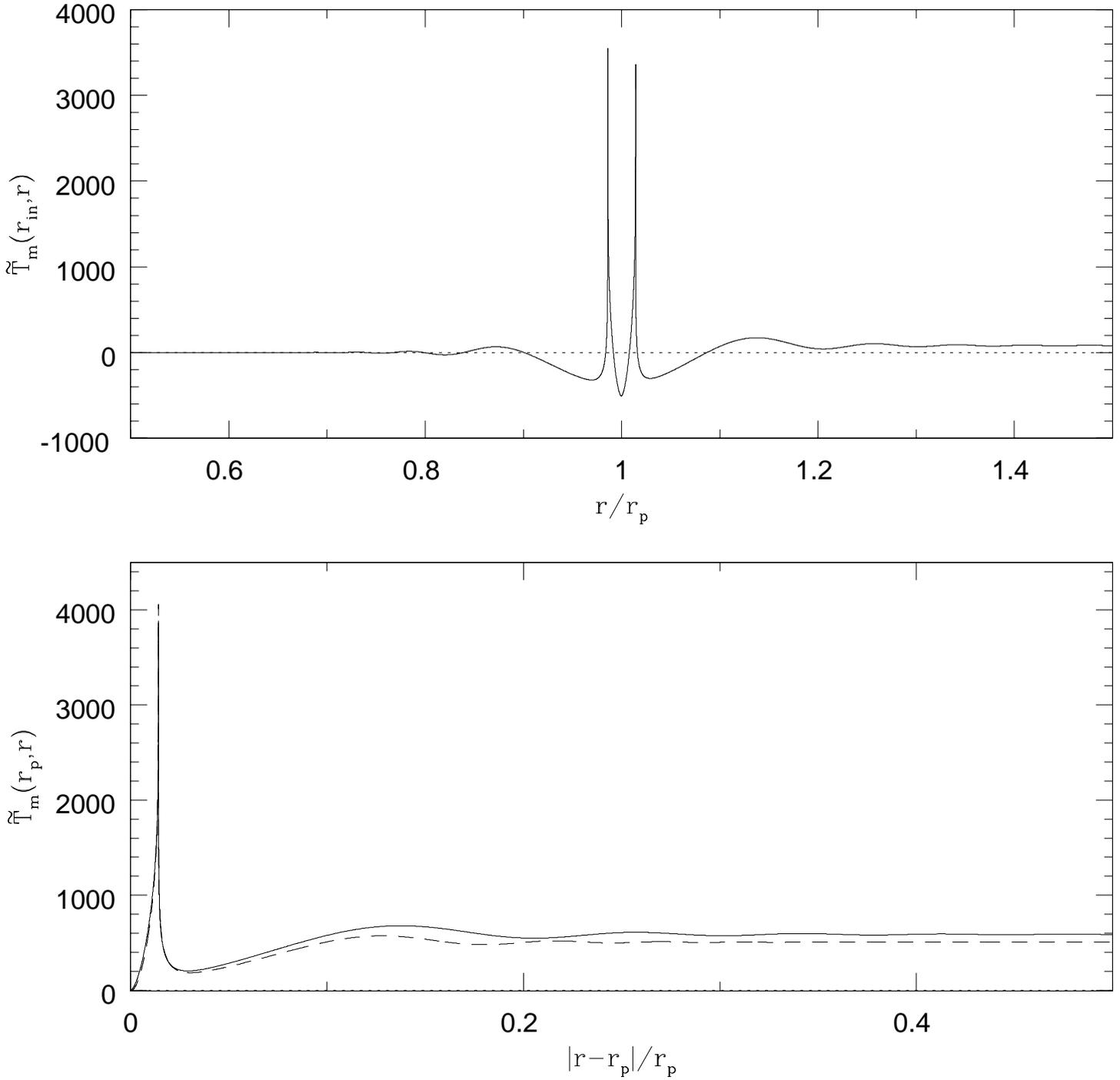} }
\caption[]{Torque $\tilde{T}_m$ exerted by the planet on the disc
for the same parameters as in fig.~\ref{fig3}.  The upper panel shows
the torque exerted between the radii $r_{in}$ and $r$ versus $r/r_p$.
The lower panel shows the torque exerted between the radii $r_p$ and
$r$ versus $|r-r_p|/r_p$, for both $r>r_p$ ({\em solid line}) and
$r<r_p$ ({\em dashed line}). Here the total torque on the disc is
positive, like in the nonmagnetic case. }
\label{fig5}
\end{figure}

We first consider the case $d_1=0$, $c/ \left( r_p \Omega_p \right) =
0.03$ ($c_1=0$), $b_1=2$, $ \beta = {\rm const} = 1$ and $m=10$.  In
figures~\ref{fig3}, \ref{fig4} and~\ref{fig5} we plot $\tilde{W}'_m$,
$\tilde{T}_{m,D}$ and $\tilde{T}_m$ versus radius.  Note that the
rapid variation of ${\rm Re}(\tilde{W}'_m)$ at $r=r_p$ is due to the
rapid variation of the potential there, not to a singularity of the
homogeneous differential equation.  We have matched $W'_m$ obtained
numerically with the analytical expression~(\ref{Wmm}) in order to
compute the constants $C_\epsilon$.  We find that ${\rm Re}(C_{+1})>0$
whereas ${\rm Re}(C_{-1})<0$.  According to
equation~(\ref{torqueres}), the point--like torque exerted at the
outer magnetic resonance is then negative whereas that exerted at the
inner magnetic resonance is positive.  This is in agreement with the
curves shown in figure~\ref{fig4} around resonances.
Figure~\ref{fig5} shows the total torque computed using the numerical
form of the torque density.  As a check, we have also calculated
$\tilde{T}_m$ by using the analytical expression~(\ref{torquedensM})
for the torque density around the resonances (represented by the
dashed lines in the lower panels of fig.~\ref{fig4}).  In general the
two calculations are in good agreement with each other, although in
some cases the discrepancy may reach about 50\% (the sign of the
torque does not change however).

\begin{figure}
\centerline{ \epsfig{file=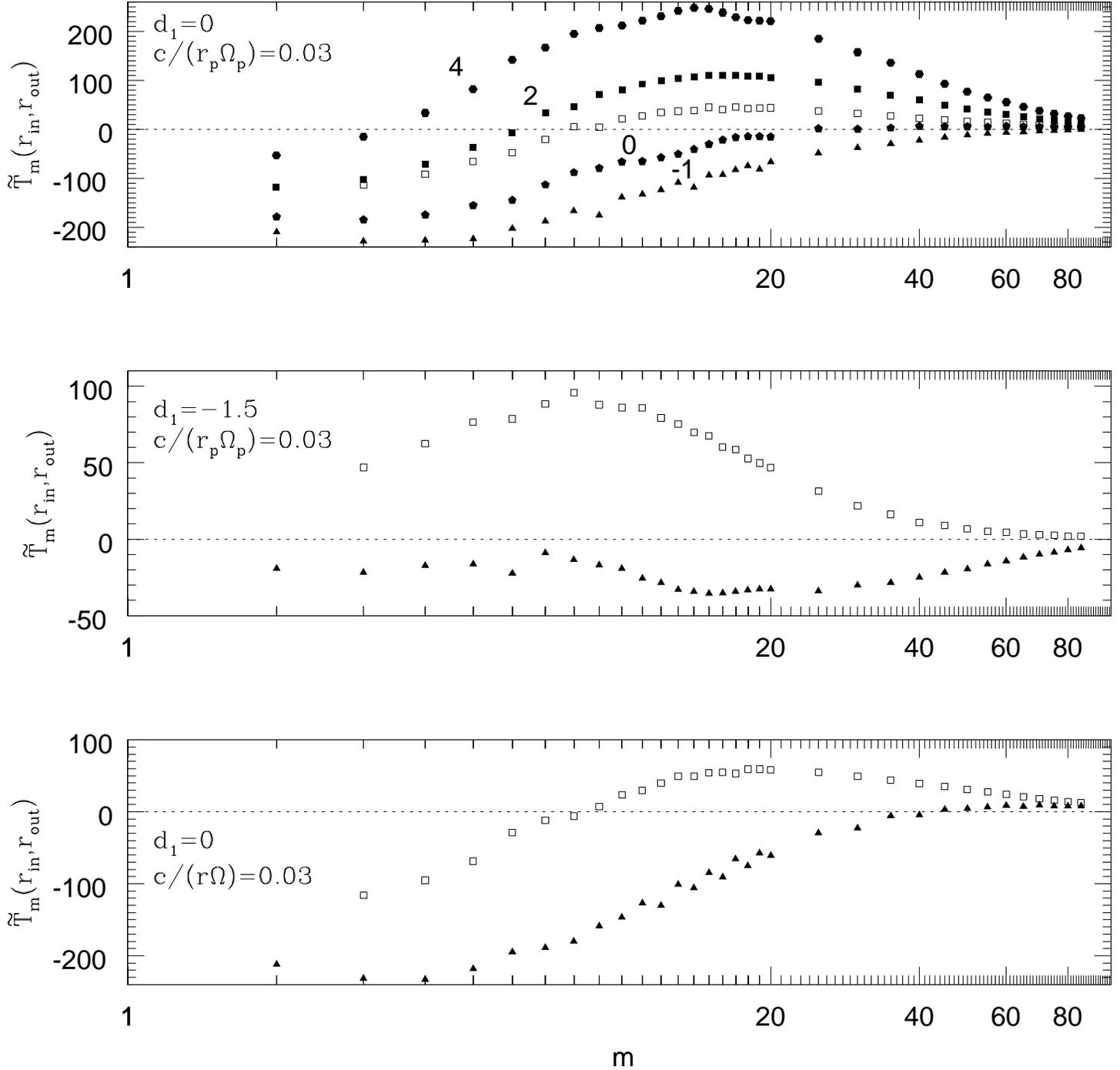} }
\caption[]{Total torque $\tilde{T}_m(r_{in}, r_{out})$ versus $m$ for
$d_1=0$ ({\em upper and lower panels}), and -1.5 ({\em middle panel})
and for $c/(r_p \Omega_p)=0.03$, i.e.  $c_1=0$ ({\em upper and middle
panels}) and $c/(r \Omega)=0.03$, i.e.  $c_1=-0.5$ ({\em lower
panel}). The open squares correspond to $\beta=\infty$, i.e. $B=0$.
The filled symbols correspond to $\beta(r_p)=1$ and $b_1=-1$ ({\em
triangles}), 0 ({\em pentagons }), 2 ({\em squares}) and 4 ({\em
hexagons}), as indicated by the label on the curves.  In the upper
panel, an estimate of the cumulative torque gives $\tilde{T}=-3975$,
-1300, 4160 and 9148 for $B \ne 0$ and $b_1=-1$, 0, 2 and 4,
respectively, and $\tilde{T}=1362$ for $B=0$.  In the middle panel,
$\tilde{T}=-1768$ for $b_1=-1$ and $\tilde{T}=2051$ for $B=0$.  In the
lower panel, $\tilde{T}=-2680$ for $b_1=-1$ and $\tilde{T}=2383$ for
$B=0$.  A negative (positive) cumulative torque corresponds to the
planet migrating outward (inward).}
\label{fig6}
\end{figure}

In figure~\ref{fig6}, we plot the total torque $\tilde{T}_m$ exerted
by the planet on the disc versus $m$ for different values of $d_1$,
$c_1$ and $b_1$ and for both $B=0$ and $B \ne 0$.  In the magnetic
case, the value of $\beta$ at $r_p$ is taken to be 1.  The cumulative
torque $\tilde{T}$ is obtained by summing up over $m$.  We get an
estimate by assuming $\tilde{T}_m$ is constant in the intervals $20
\le m \le 24$, $25 \le m \le 29$ etc... and $\tilde{T}_m=0$ for
$m>85$.

The values computed for $B=0$ are in very good agreement with KP93.
In this case, the cumulative torque exerted by the planet on the
disc is positive, which means that the planet loses angular momentum
and migrates inward.  When the magnetic field is non zero though,
$\tilde{T}$ may become negative, with a subsequent outward planet
migration.  This occurs when $b_1$ is negative enough, i.e. when
$\beta$ increases fast enough with radius or, equivalently, when the
magnetic field decreases fast enough with radius.  For instance, when
$d_1=c_1=0$, $\tilde{T}<0$ for $b_1=-1$, i.e. $\beta \propto r^3$ or,
equivalently, $\langle B^2 \rangle \propto r^{-3}$, and for $b_1=0$,
i.e. $\beta \propto r^2$ or, equivalently, $\langle B^2 \rangle
\propto r^{-2}$.  Note that this variation needs only be {\em local}.
When $b_1$ goes from positive to negative values, the cumulative
torque, initially positive, then becomes negative passing through
zero.  Therefore migration is slowed down and then reversed.

\begin{figure}
\centerline{ \epsfig{file=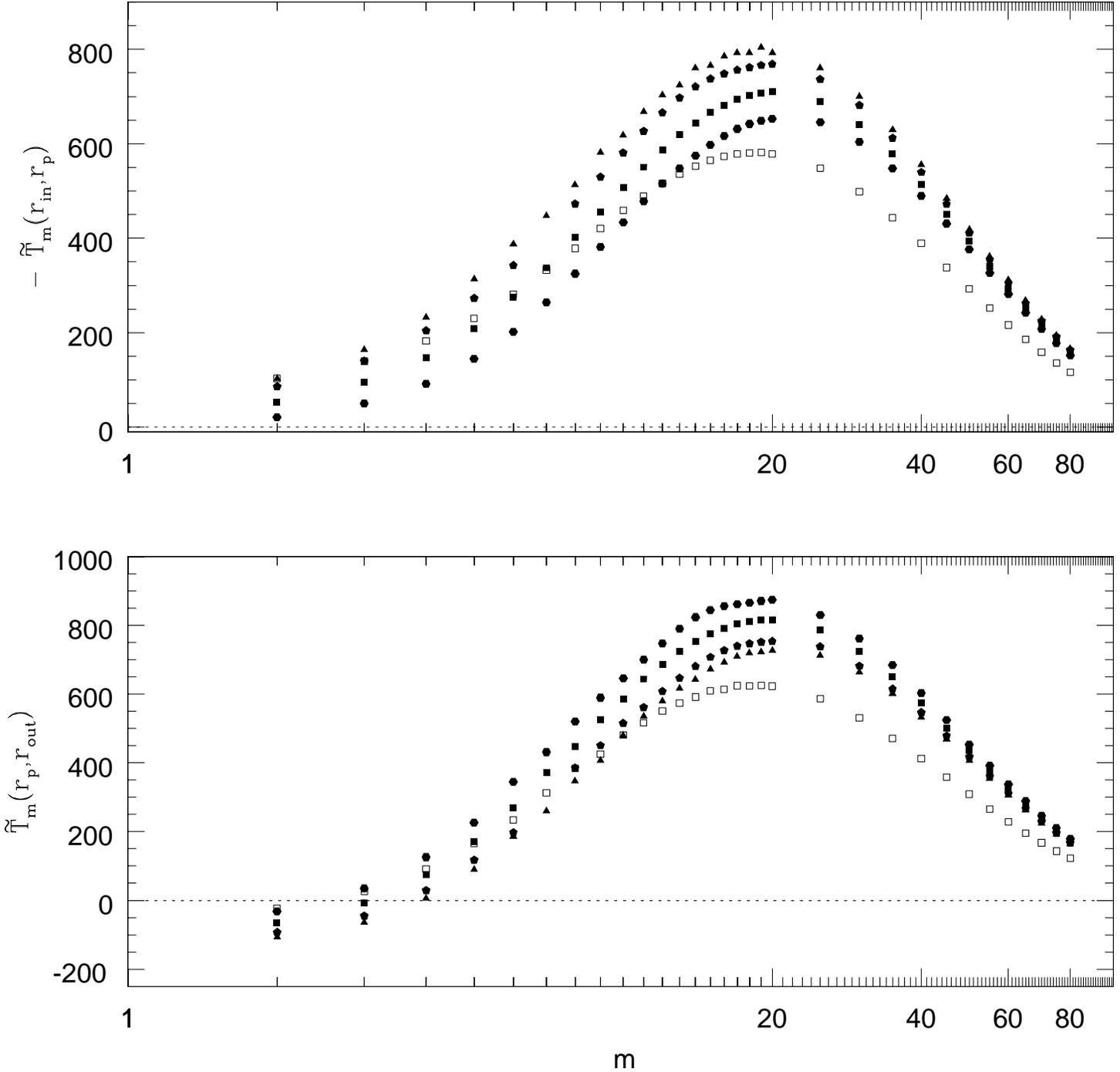} }
\caption[]{{\em Upper panel}: minus the torque exterted by the planet
on the disc inside its orbit, -$\tilde{T}_m(r_{in}, r_{p})$, versus
$m$. {\em Lower panel}: torque exterted by the planet on the disc
outside its orbit, $\tilde{T}_m(r_{p}, r_{out})$, versus $m$.  The
parameters are the same as in the upper panel of figure~\ref{fig6}:
$d_1=0$ and $c/(r_p \Omega_p)=0.03$, i.e.  $c_1=0$.  Like in
figure~\ref{fig6}, the open squares correspond to $\beta=\infty$,
i.e. $B=0$, and the filled symbols correspond to $\beta(r_p)=1$ and
$b_1=-1$ ({\em triangles}), 0 ({\em pentagons }), 2 ({\em squares})
and 4 ({\em hexagons}). An estimate of the cumulative torque for $m
\le 80$ gives $[\tilde{T}(r_{in}, r_{p}); \tilde{T}(r_{p}, r_{out})] =
(-39531;35560), (-37727;36521), (-35442;39535), (-32856;42161)$ for $B
\ne 0$ and $b_1=-1$, 0, 2 and 4, respectively, and $(-28278;29486)$
for $B=0$.  The one--sided torque has a larger amplitude when a
magnetic field is present.  }
\label{fig6b}
\end{figure}

In figure~\ref{fig6b}, we plot the one--sided torque exterted by the
planet on the disc (i.e. the torque exterted inside and outside its
orbit) versus $m$ for $d_1=0$ and $c/(r_p \Omega_p)=0.03$, i.e.
$c_1=0$.  These plots show that the one--sided torque has a larger
amplitude when a magnetic field is present.

\begin{figure}
\centerline{ \epsfig{file=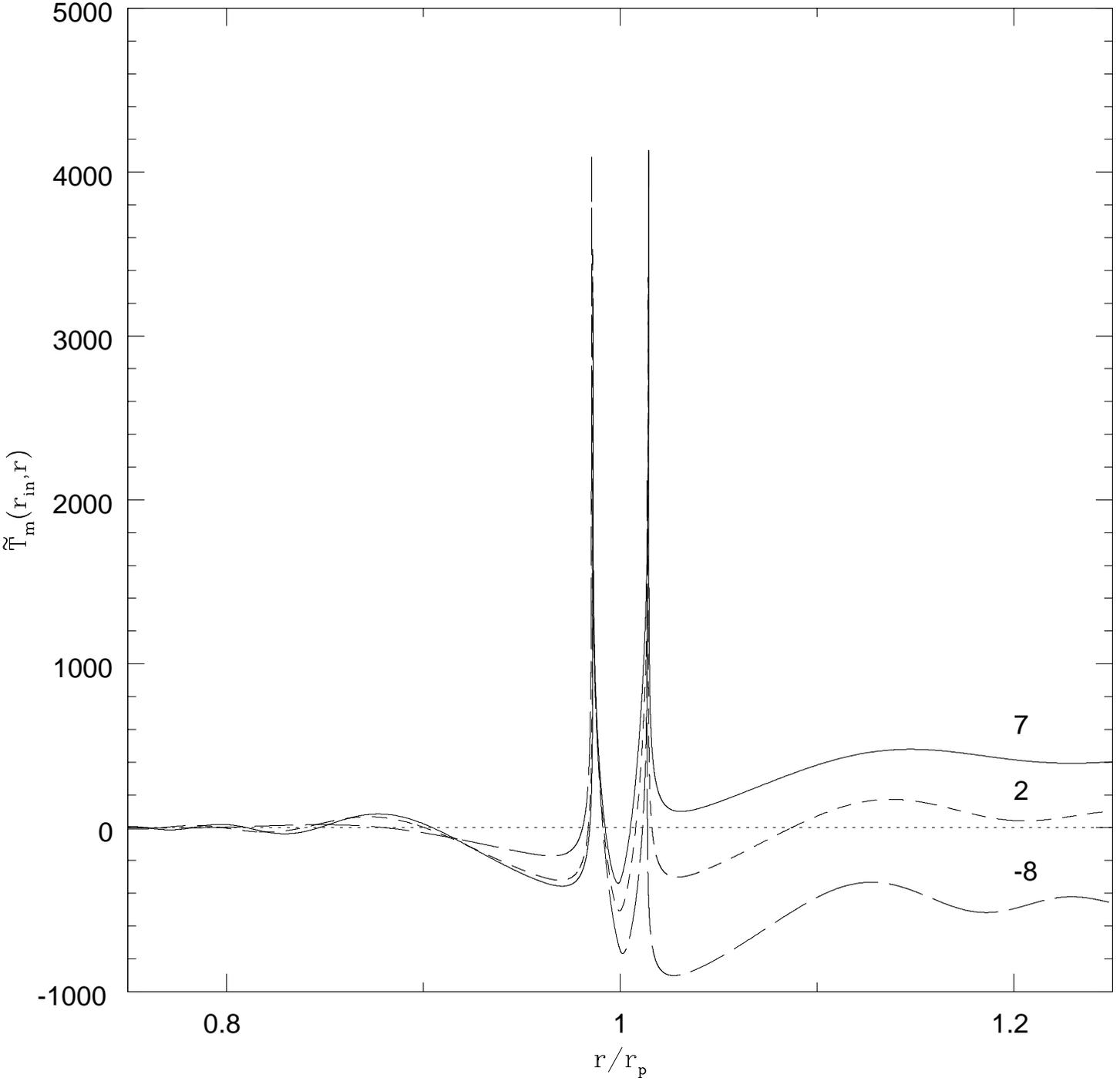} }
\caption[]{Torque $\tilde{T}_m(r_{in}, r)$ exerted between $r_{in}$
and $r$ versus $r/r_p$ for $m=10$, $d_1=0$, $c/(r_p \Omega_p)=0.03$
($c_1=0$), $\beta(r_p)=1$ and $b_1=7$ ({\em solid line}), 2 ({\em
short--dashed line}) and -8 ({\em long--dashed line}).  As $b_1$
decreases, the torque exerted inside the Lindblad resonances becomes
increasingly dominant and negative.}
\label{fig7}
\end{figure}

In figure~\ref{fig7} we plot the torque exerted between $r_{in}$
and $r$ versus $r/r_p$ for $m=10$, $d_1=c_1=0$, $\beta(r_p)=1$ and for
$b_1=7, 2$ and $-8$.  This shows that the total torque becomes
negative when $b_1$ is negative due to the contribution from the
region inside the Lindblad resonances.

In table~\ref{tab1} we have listed the value of the torque exerted
on different radii intervals in the disc for $m=10$, $d_1=0$,
$c/(r_p \Omega_p)=0.03$ ($c_1=0$), $\beta(r_p)=1$ and different values
of $b_1$.  The torque exerted between the inner Lindblad resonance
$r_{ILR}$ and corotation $r_p$ is negative, whereas that exerted
between corotation and the outer Lindblad resonance $r_{OLR}$ is
positive.  (Note that for $m=10$ the Lindblad resonances co\"{\i}ncide
with the outermost turning points).  We have found above that the
point--like torque exerted at the inner magnetic resonance is positive
whereas that exerted at the outer resonance is negative.  Therefore,
the torque is not dominated by the point--like contribution at the
magnetic resonances, but is determined by the whole region around
these resonances, where waves propagate for $m<m_{crit}$ (see also the
discussion in section~\ref{sec:torquereson}).  When $b_1$ decreases,
$\beta$ increases outward faster, so that the magnetic field becomes
less and less important outside the planet's orbit.  Therefore the
torque exerted between $r_p$ and $r_{OLR}$, which is positive and
tends to be negligible in the absence of a magnetic field, becomes
less and less important.  In contrast, the magnetic field becomes more
important inside the orbit, where it gives rise to a negative torque.
The net effect of decreasing $b_1$ is then to decrease the total
torque.  If the magnetic field is strong enough, the total torque
exerted on the disc is dominated by the region inside the Lindblad
resonances (which is closer to the planet than the regions outside
these resonances), and it can then become negative for small enough
values of $b_1$.

\begin{figure}
\centerline{ \epsfig{file=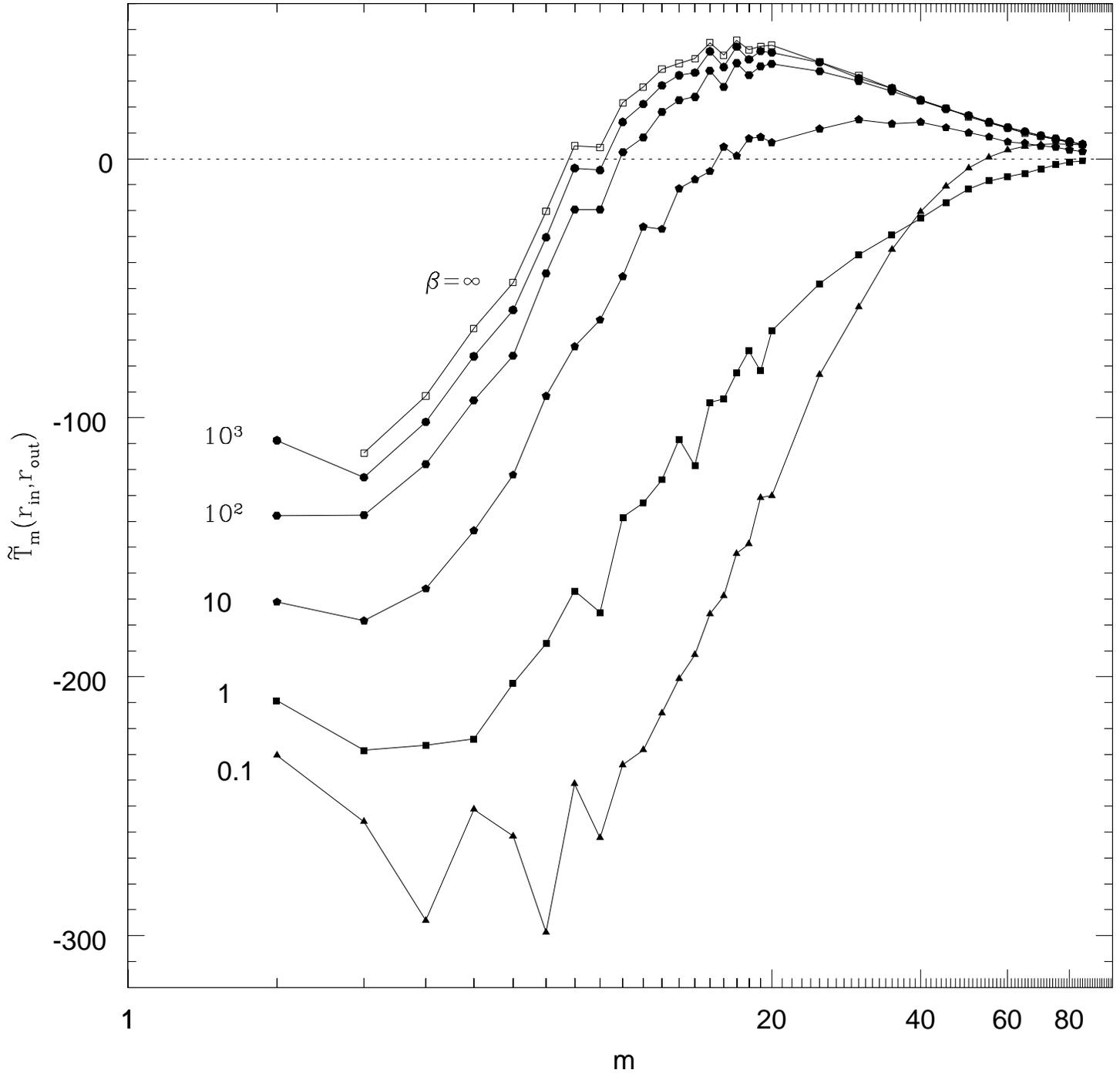} }
\caption[]{Total torque $\tilde{T}_m(r_{in}, r_{out})$ versus $m$ for
$d_1=0$ and $c/(r_p \Omega_p)=0.03$ ($c_1=0$).  The open squares
correspond to $\beta=\infty$, i.e. $B=0$.  The filled symbols
correspond to $b_1=-1$ and $\beta(r_p)=0.1$ ({\em triangles}), 1 ({\em
squares}), 10 ({\em pentagons}) $10^{2}$ ({\em hexagons}) and
$10^{3}$ ({\em septagons}).  An estimate of the cumulative torque
gives $\tilde{T}=1362$, 1129, 855, -508, -3975 and -5483 for
$\beta(r_p)=\infty$, $10^{3}$, $10^{2}$, 10, 1 and 0.1, respectively.
This shows that $\tilde{T}$ is significantly reduced for values of
$\beta$ as large as $10^2$.  }
\label{fig8}
\end{figure}

Figure~\ref{fig8} shows the total torque exerted by the planet on
the disc versus $m$ for $d_1=0$, $c/(r_p \Omega_p)=0.03$ ($c_1=0$),
$b_1=-1$, which correspond to $\beta \propto r^3$, and different values
of $\beta(r_p)$.  As we can see, the cumulative torque $\tilde{T}$ is
significantly decreased for values of $\beta(r_p)$ as large as
$10^{2}$ and becomes negative for values of $\beta(r_p)$ between
$10^{2}$ and 10.

The cumulative torque corresponding to $\beta(r_p)=10^{2}$ becomes
negative if, for the same parameters as above, $d_1$ is taken to be at
least as large as unity, i.e. if $\Sigma$ increases with radius at
least as fast as linearly.  This is because when $d_1$ increases
$\beta$ increases faster with radius, which favours the negative inner
torque exerted near corotation.  This effect can be seen in
figure~\ref{fig6} by comparing the upper and middle panels, which show
that the total torque corresponding to $d_1=-1.5$ is larger than that
corresponding to $d_1=0$ for $b_1=-1$.  Note that $d_1>0$ tends to
favour the outer Lindblad resonance compared to the inner one, which
results in a larger Lindblad torque (Ward 1986, 1997).  The fact that
the total torque actually decreases when $d_1$ increases means that
the variations of $\Sigma$ affect more significantly the torque
exerted around the magnetic resonances than the Lindblad torque.
(Note that the variations of the torque displayed in figure~\ref{fig6}
when $d_1$ varies in the nonmagnetic case are mainly due to changes
in the corotation torque).

An increase of $c_1$ (corresponding to $c$ or, equivalently, $H/r$,
increasing faster with radius) also results in a smaller total torque,
as can be seen by comparing the upper and lower panels of
figure~\ref{fig6}.  When $c_1$ gets larger, $\beta$ increases faster
with radius, which again lowers the torque exerted around the magnetic
resonances.  Also the Lindblad torque is reduced when $c_1$ gets
larger (Ward 1986, 1997).

\begin{figure}
\centerline{ \epsfig{file=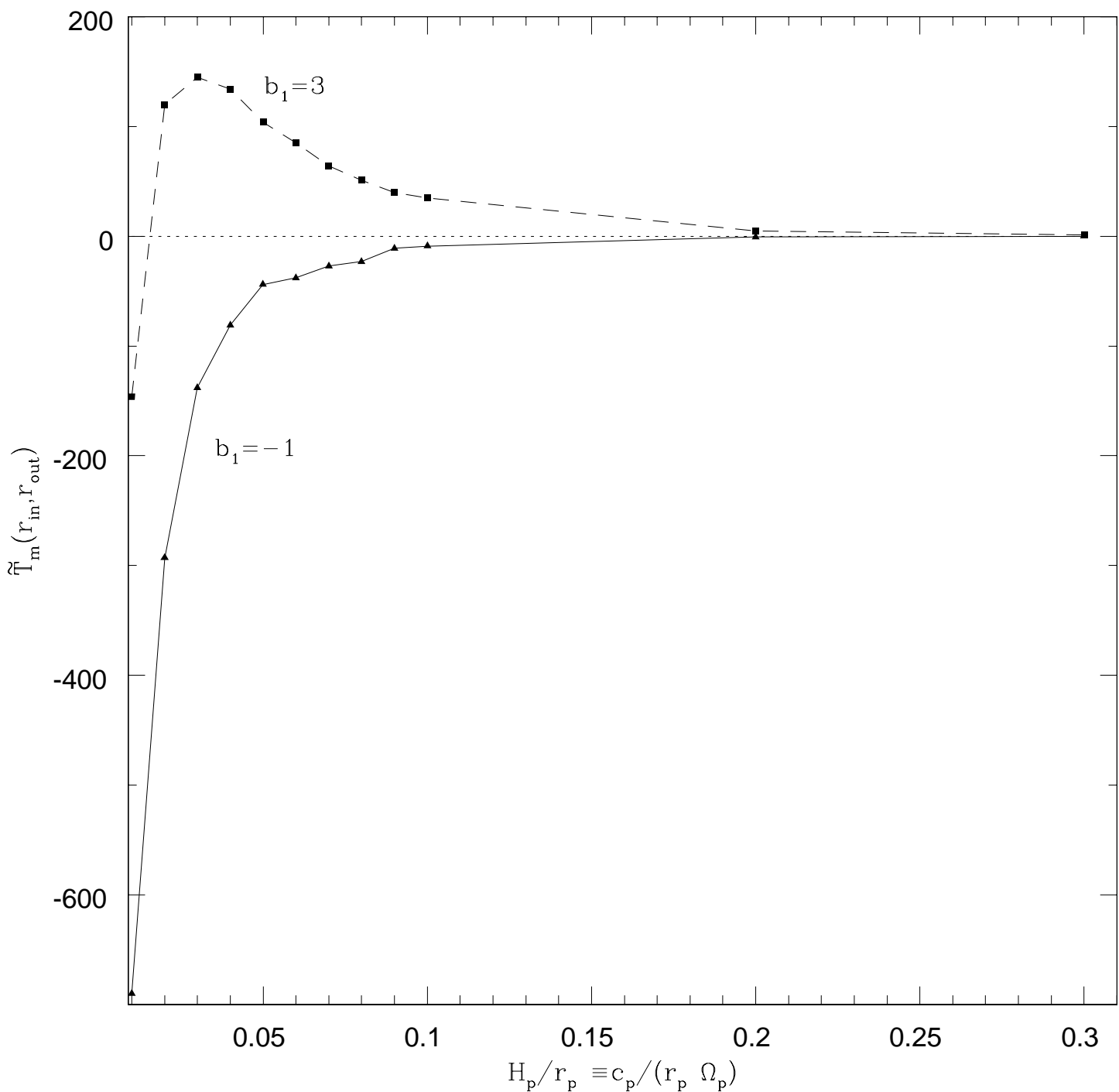} }
\caption[]{Total torque $\tilde{T}_m(r_{in}, r_{out})$ versus $H_p/r_p
\equiv c_p/(r_p \Omega_p)$ (where the subscript 'p' indicates that the
quantities are taken at $r=r_p$) for $m=10$, $d_1=0$, $c_1=0$,
$\beta(r_p)=1$ and $b_1=-1$ ({\em solid line}) and 3 ({\em dashed
line}).  The torque tends to zero as $H/r$ increases.  }
\label{fig9}
\end{figure}

Figure~\ref{fig9} shows the effect of varying the value of $H/r \equiv
c/(r \Omega)$ at $r=r_p$ for $m=10$, $d_1=0$, $c_1=0$, $\beta(r_p)=1$
and $b_1=-1$ and 3.  The total torque exerted on the disc tends to
zero as $H/r$ increases.  This is because the resonances are further
away from the planet when $H/r$ is larger, so that the coupling
between the perturbation and the tidal potential is weaker.  For
$b_1=-1$, the torque is always negative and decreases as $H/r$
decreases.  For $b_1=3$, the torque is positive and maximum for some
value of $H/r$ ($\sim 0.03$).  It stays positive (and becomes smaller)
as $H/r$ increases, whereas it becomes negative for smaller values of
$H/r$.  This is because even though $\beta$ increases with radius in
that case, the region between the inner Lindblad resonance and
corotation contributes to a larger magnitude of the torque than the
region between corotation and the outer Lindblad resonances.  This is
also the case for $H/r=0.03$ and $b_1=1$, as shown in
table~\ref{tab1}.

Note that what determines the sign of the torque is the gradient of
$\beta$, not of the magnetic field itself, as only $\beta$ enters into
equation~(\ref{EQ2}).

\section{Discussion and conclusion}
\label{sec:discussion}

We have calculated the linear torque exerted by a planet on a circular
orbit on a disc containing a toroidal magnetic field.  In contrast to
the nonmagnetic case, there is no singularity at the corotation
radius, where the frequency of the perturbation matches the orbital
frequency.  However, there are two new singularities on both sides of
corotation, where the frequency of the perturbation in a frame
rotating with the fluid matches that of a slow MHD wave propagating
along the field line.  These so--called {\em magnetic resonances} are
closer to the planet than the Lindblad resonances.  In addition, on
each side of the planet, there are two turning points located beyond
the magnetic resonance.  One of them co\"{\i}ncides with the Lindblad
resonance.  For values of $m$ larger than some critical value $m_{\rm
crit}$, there is a third turning point between the magnetic resonance
and corotation, on each side of the planet.  Like in the nonmagnetic
case, waves propagate outside the outermost turning points.  But here
they also propagate inside the intermediate turning points when
$m<m_{\rm crit}$, or between the intermediate and innermost turning
points when $m>m_{\rm crit}$ (see figure~\ref{fig1}).  The singular
modes excited at the magnetic resonances can therefore propagate.

There is a significant torque exerted on the region of the disc inside
the outermost turning points (which co\"{\i}ncide with the Lindblad
resonances for the values of $m$ of interest).  Like the Lindblad
torque, this torque is negative inside the planet's orbit and positive
outside its orbit.  The whole region around corotation, not just a
narrow zone around the magnetic resonances, contribute to this torque.
In other words, the magnetic resonances contribute to a global, not a
point-like, torque.  Since these resonances are closer to the planet
than the Lindblad resonances, they couple more strongly to the tidal
potential.  Therefore, the torque exerted around the magnetic
resonances dominate over the Lindblad torque if the magnetic field is
large enough.  If in addition $\beta \equiv c^2/v_A^2$ increases fast
enough with radius, the outer magnetic resonance becomes less
important (it disappears altogether when there is no magnetic field
outside the planet's orbit) and the total torque is then negative,
dominated by the inner magnetic resonance.  This corresponds to a
positive torque on the planet, which leads to outward migration.

The amount by which $\beta$ has to increase outward for the total
torque exerted on the disc to be negative depends mainly on the
magnitude of $\beta$.  We have found that for $\beta =1$ at
corotation, the cumulative torque (obtained by summing up the
contributions from all the values of $m$) exerted on the disc is
negative when $\beta$ increases at least as fast as $r^2$.  If $\beta
\propto r^3$, the cumulative torque becomes negative for values of
$\beta(r_p)$ between $10^2$ and 10, whereas it is negative for
$\beta(r_p)=10^2$ if $\beta \propto r^4$.

The migration timescales that correspond to the torques calculated
above are rather short.  The orbital decay timescale of a planet of
mass $M_p$ at radius $r_p$ is $\tau = M_p r_p^2 \Omega_p / |T|$, where
$T$ is the cumulative torque exerted by the planet on the disc.
Remembering that $T \propto (M_p/M_\ast)^2$, we get:

\begin{equation}
\tau ({\rm yr}) = 4.3 \times 10^9 \left( \frac{M_p}{{\rm M_\oplus}}
\right)^{-1} \left( \frac{ \Sigma_p }{100 \; {\rm g \; cm^{-2}}}
\right)^{-1} \left( \frac{r_p}{1 \; {\rm au}} \right)^{-1/2}
\tilde{T}^{-1}.
\end{equation}

\noindent In a standard disc model, $\Sigma \sim
100$--$10^3$~g~cm$^{-2}$ at 1~au (see, for instance, Papaloizou \&
Terquem~1999).  Therefore, $\tau \sim 10^5$--$10^6$~yr for a one earth
mass planet at 1~au in a nonmagnetic disc, as $\tilde{T} \sim 10^3$ in
that case (see fig.~\ref{fig6}).  This is in agreement with Ward
(1986, 1997).  In a magnetic disc, $|\tilde{T}|$ may become larger,
leading to an even shorter migration timescale (note that given the
limited accuracy of the numerical scheme used, as indicated in
section~\ref{sec:caseBne0}, only orders of magnitude for the migration
timescale are obtained here).  However, it is important to keep in
mind that these timescales are {\em local}.  Once the planet migrates
outward out of the region where $\beta$ increases with radius, it may
enter a region where $\beta$ behaves differently and then resume
inward migration for instance.  Such a situation would be expected to
occur in a turbulent magnetized disc in which the large scale field
structure changes sufficiently slowly.

Unless the magnetic field is above equipartition, it is unstable by
the magnetorotational instability (Balbus \& Hawley 1991, 1998 and
references therein), the saturated nonlinear outcome of which is MHD
turbulence.  In a turbulent magnetized disc, the main component of the
field is toroidal, as shown by numerical simulations (Hawley, Gammie
\& Balbus 1995; Brandenburg et al. 1995).  Global simulations have
also shown that the turbulence saturates at a level corresponding to
$\beta \sim 100$ if there is no mean flux or to lower values of
$\beta$ if there is a mean flux (Hawley 2001; Steinacker \& Papaloizou
2002).  Also, substantial spatial inhomogeneities are created by
radial variations of the Maxwell stress (Hawley 2001; Steinacker \&
Papaloizou 2002), so that the field may display significant gradients.
On the basis of the work presented here, we are led to speculate that
in such a disc the planet would suffer alternately inward and outward
migration, or even no migration at all.  It would then oscillates back
and forth in some region of the disc, or undergo some kind of
diffusive migration, either outward or inward depending on the field
gradients encountered.  Note however that in a turbulent disc the
magnetic field may vary locally on timescales shorter than the
timescales needed to establish the type of tidal response assumed in
this paper.

It has been pointed out that protoplanetary discs may be ionized
enough for the magnetic field to couple to the matter only in their
innermost and outermost parts (Gammie 1996; Fromang, Terquem \& Balbus
2002).  A planet forming at around one astronomical unit, where the
ionization fraction is very low, would then migrate inward on the
timescale calculated by Ward (1986, 1997) until it reaches smaller
radii where the field is coupled to the matter.  At this point, there
would be a magnetic field inside the planet's orbit but not outside
its orbit.  The torque on the planet would then reverse, and outward
migration would occur.  However, as soon as the planet would re--enter
the nonmagnetic region, inward migration would resume.  Hence, the
planet would stall at the border between the magnetic and nonmagnetic
regions.

Inward migration in a magnetized disc may then either be very
significantly slowed down, occur only on limited scales, or not occur
at all.  The planet would then be able to grow to become a terrestrial
planet or the core of a giant planet.

When the planet becomes massive enough (about 10 earth masses), the
interaction with the disc becomes nonlinear, with a gap being opened
up around the planet's orbit.  Because a rather strong torque is
exerted in the vicinity of the planet in the presence of a magnetic
field, a gap should open up more easily and be 'cleaner' in a
turbulent magnetic disc than in a similar laminar viscous disc.  This
seems to be in agreement with the numerical simulations of a giant
planet in a turbulent disc performed by Nelson \& Papaloizou (2002).
Once the gap is completely cleared out, the effect of the magnetic
resonances disappear as they are inside the edges of the gap (which
co\"{\i}ncide with the Lindblad resonances corresponding to $m \sim
r/H$).  The magnetic field may however still have an effect on the
disc--planet interaction because of the turbulence it produces (Nelson
\& Papaloizou 2002).

We have focused here on a planet with a circular orbit.  A planet on
an eccentric orbit in a nonmagnetic disc usually suffers eccentricity
damping, as in that case the corotation resonances, which damp the
eccentricity, dominate over the Lindblad resonances, which excite it
(Goldreich \& Tremaine 1980; Papaloizou, Nelson \& Masset 2001).
This picture may change once a magnetic field is introduced.  The
effect of a magnetic field on the eccentricity of a planet will be
studied in another paper.

\section*{acknowledgement}

It is a pleasure to thank Steven Balbus and John Papaloizou for their
advice, encouragement, and many valuable suggestions and discussions.
I am grateful to John Papaloizou for his comments on an early draft of
this paper which led to significant improvements.

\newpage


\begin{table}
\begin{center}
\begin{tabular}{llllllll} \hline \hline
$b_1$ & $\tilde{T}_m(r_{in},r_{ILR})$  & $\tilde{T}_m(r_{ILR},r_p)$ & 
$\tilde{T}_m(r_p,r_{OLR})$ & $\tilde{T}_m(r_{OLR},r_{out})$ & 
Total torque $\tilde{T}_m$ \\ 
\hline \\
8   & -172 & -110 & 624 & 194 & 536 \\
7   & -171 & -153 & 553 & 186 & 415 \\
3   & -165 & -305 & 408 & 208 & 146 \\
2   & -163 & -344 & 377 & 209 & 79 \\
1   & -160 & -385 & 333 & 211 & -1 \\
0   & -158 & -423 & 298 & 217 & -66 \\
-1  & -154 & -463 & 263 & 217 & -137 \\
-3  & -155 & -570 & 184 & 217 & -324 \\
-5  & -145 & -641 & 129 & 225 & -432 \\
-8  & -105 & -645 & 69  & 225 & -456 \\
    &      &      &     &     & \\
    & -126 & -333 & 310 & 170 & 21 \\
\hline
\hline
\end{tabular}
\end{center}
\caption{ \label{tab1} Torque $\tilde{T}_m(r_1,r_2)$ exerted by the
planet on the disc between the radii $r_1$ and $r_2$ in units
$\Sigma_p r_p^4 \Omega_p^2$ for different values of $b_1$ and for
$m=10$, $d_1=0$, $c/ (r_p \Omega_p) =0.03$ ($c_1=0$) and
$\beta(r_p)=1$.  The radii $r_1$ and $r_2$ are either the disc inner
radius $r_{in}$, the inner Lindblad resonance $r_{ILR}$, the
corotation radius $r_p$, the outer Lindblad resonance $r_{OLR}$ or the
disc outer radius $r_{out}$.  For comparison, the last line shows the
torques in the nonmagnetic case.  As $b_1$ decreases, the torque
exerted inside the Lindblad resonances becomes negative and
dominant.}
\end{table}

\newpage

\end{document}